%% file: main.tex
\documentclass[reprint,amsmath,amssymb,floatfix,aps]{revtex4-2}
\input{config/includes.tex}

\input{config/defs.tex}

\begin{document}

\title{Information Flow in Non-Unitary Quantum Cellular Automata}

\author{Elisabeth Wagner}
\email{elisabeth.wagner@students.mq.edu.au}
\author{Ramil Nigmatullin}
\author{Alexei Gilchrist}
\author{Gavin K.~Brennen}
 \affiliation{Centre for Engineered Quantum Systems, Dept.~of Physics \& Astronomy, Macquarie University, 2109 NSW, Australia}
\date{\today}

\begin{abstract}
    The information flow in a quantum system is a fundamental feature of its dynamics. 
    An important class of dynamics are quantum cellular automata (QCA),
    systems with discrete updates invariant in time and space,
    for which an index theory has been proposed for the quantification of the net flow of quantum information across a boundary.
    While the index is rigid in the sense of begin invariant under finite-depth local circuits,
    it is not defined when the system is coupled to an environment, i.e.~for non-unitary time evolution of open quantum systems. 
    We propose a new measure of information flow for non-unitary QCA denoted the information current which is not rigid, but can be computed locally based on the matrix-product operator representation of the map.

    %


\end{abstract}

\maketitle

\section{Introduction}

The essential physical principles of causality and conservation of information impose strong constraints on the time evolution of physical systems. 
In particular, in the simplest setting where space and time are discrete and causality is preserved, 
quantum many-body systems can be described by quantum cellular automata (QCA) \cite{Farrelly_2020,Arrighi_2019,Schumacher_2004},
which are systems with discrete variables evolving under a local update rule (in analogy with classical cellular automata).
Despite these seemingly crude approximations for realistic many-body dynamics, QCA provide useful models to study different aspects of non-equilibrium physics: local quantum circuits, a subclass of QCA, have recently received significant attention in connection to questions related to quantum chaos and information scrambling \cite{Xu:2020va,PhysRevX.8.031058,PhysRevLett.123.210601,PhysRevB.98.134204,PhysRevResearch.2.033032,PhysRevX.8.021013,PhysRevX.8.031057,PhysRevX.7.031016,PhysRevB.102.064305,PhysRevX.8.041019,PhysRevLett.123.210603,PhysRevLett.121.060601}.

In the past decade, a great deal of progress has been made in the comprehensive characterization of QCA: The so-called index theory was first introduced in \cite{index_werner} for one-dimensional systems and has recently been extended and generalized to higher dimensional systems in \cite{index_freedman,haah2018nontrivial,freedman2019group}.
In one dimension, the index (or GNVW index according to the acronym of the authors) describes the net flow of quantum information along a chain of, say qubits; while in higher dimensions it is given by the information flow between two subsystems of the total quantum grid of logical qubits.
For unitary one-dimensional systems it takes the form of a positive rational fraction, $\tx{ind} \in \mathbb Q_+$, which can be interpreted as the ratio of the number of orthonormal states transferred to the right divided by the number transferred to the left after each discrete time step.
Besides its fundamental interest, the index theory has turned out to have practical implications, allowing, for instance, for a classification of 2D Floquet phases exhibiting bulk many-body localization (MBL) \cite{Arrighi_2019,PhysRevX.6.041070,Zhang_2021,PhysRevB.95.155126,PhysRevLett.118.115301,liu2020gauging,PhysRevB.99.085115,PhysRevB.98.054309}, where the index serves as a topological invariant that measures the chirality of quantum information flow.
Formally, the GNVW index has been defined originally in terms of abstract observable algebras \cite{index_werner} which were later argued to be  
``lacking an immediate physical interpretation'' \cite{gong2020topological}
and to be ``both physically opaque and not amenable to experimental measurement'' \cite{PhysRevB.98.054309}.

Subsequently, an equivalent definition of the index has been found
by taking the entanglement of the “vectorized” evolution operator, or operator-space entanglement entropy, into consideration. The R\'enyi-$\alpha$ entropy has been shown to be an appropriate alternative measure, as it can be computed locally and closely reflects the intuitive interpretation of the index in terms of quantum information flow.
Using this quantity, any sub-linear entanglement-growth behavior in nontrivial QCA could be ruled out, and a lower bound on quantum chaos has been defined for any R\'enyi-$\alpha$ entropy of the evolution operator \cite{gong2020topological}.

The original GNVW index has been rediscovered by taking the \Renyitwo entropy into account --- a quantity that can be measured directly 
using existing “SWAP”-based many-body quantum interference setups with e.g.~hard-core bosonic ultracold atoms in a shaken optical lattice \cite{PhysRevX.6.041070}.
This formulation of the index in terms of the \Renyitwo entropy is notably equivalent to a previous derivation of the index in terms of the chiral mutual information
\cite{PhysRevB.98.054309},
as the latter can be constructed from any extensive entanglement measure, including \Renyi entropies.

Next, matrix product unitaries (MPUs) have turned out to provide a natural framework for the index theory as they have been shown to in fact be QCA and vice versa; 
i.e.~MPUs feature a causal cone, strictly propagating information over a finite distance only.
MPUs are thereby guaranteed to preserve locality by mapping local operators to local operators while at the same time all locality-preserving unitaries can be represented in a matrix product way.
The index theory implies that 
all locality-preserving 1D unitaries can be efficiently simulated by MPUs, and that
different MPU representations of the same unitary can be related through a local gauge.
The explicit computability of the GNVW index via MPUs has been demonstrated in \cite{PhysRevX.6.041070},
and has led to further physical consequences in the framework of Floquet
dynamics, where
bulk topology has been shown to enforce chaotic dynamics at the edge.

An equivalent expression of the GNVW index has been given by the ``rank-ratio'' index, which is defined as the
ratio between the ranks of the left and right singular value decompositions of the tensor representing the MPU
\cite{PhysRevB.98.245122,Ignacio_Cirac_2017}.
Based on this definition, an index theorem for generalized MPUs has been defined taking fermionic QCA into account,
where a graded canonical form has been introduced for fermionic matrix product states
\cite{Piroli_2021}.
Further, Hamiltonian evolutions on the lattice satisfying Lieb-Robinson bounds, rather than strict locality,
have been described by
approximately locality preserving unitaries (ALPUs). 
The index theory has been shown to be robust to this generalization, and has been extended to one-dimensional ALPUs classifying a wider class of natural systems with approximate causal cones only. %
A converse to the Lieb-Robinson bounds has further been achieved, where
any ALPU of index zero can be exactly generated by some time-dependent, quasi-local Hamiltonian in constant time. For the special case of finite chains with open boundaries, any unitary satisfying the Lieb-Robinson bound may be generated by such a Hamiltonian
\cite{ranard2020converse}.

While much progress on index theory has been made on unitary QCA, very little is known about discrete non-unitary systems representing more general (irreversible) physical actions \cite{PhysRevA.68.042311,Richter:1996ww,PhysRevLett.125.190402}, where a general characterization is essentially missing.

In this work, we address this question and present a measure for the net information current as an equivalence class of non-unitary QCA –– a class which is here described by one-dimensional matrix product operators (MPOs).
%
This classification will help to comprehensively characterize open quantum systems and provides a measure for the speed of net information transfer in physical systems.

To provide the reader with a first intuitive understanding of how the information flow in a quantum system can be defined, an example of a simple, one-dimensional QCA is discussed in the following.
\bfig
    \includegraphics[width=.98\columnwidth]{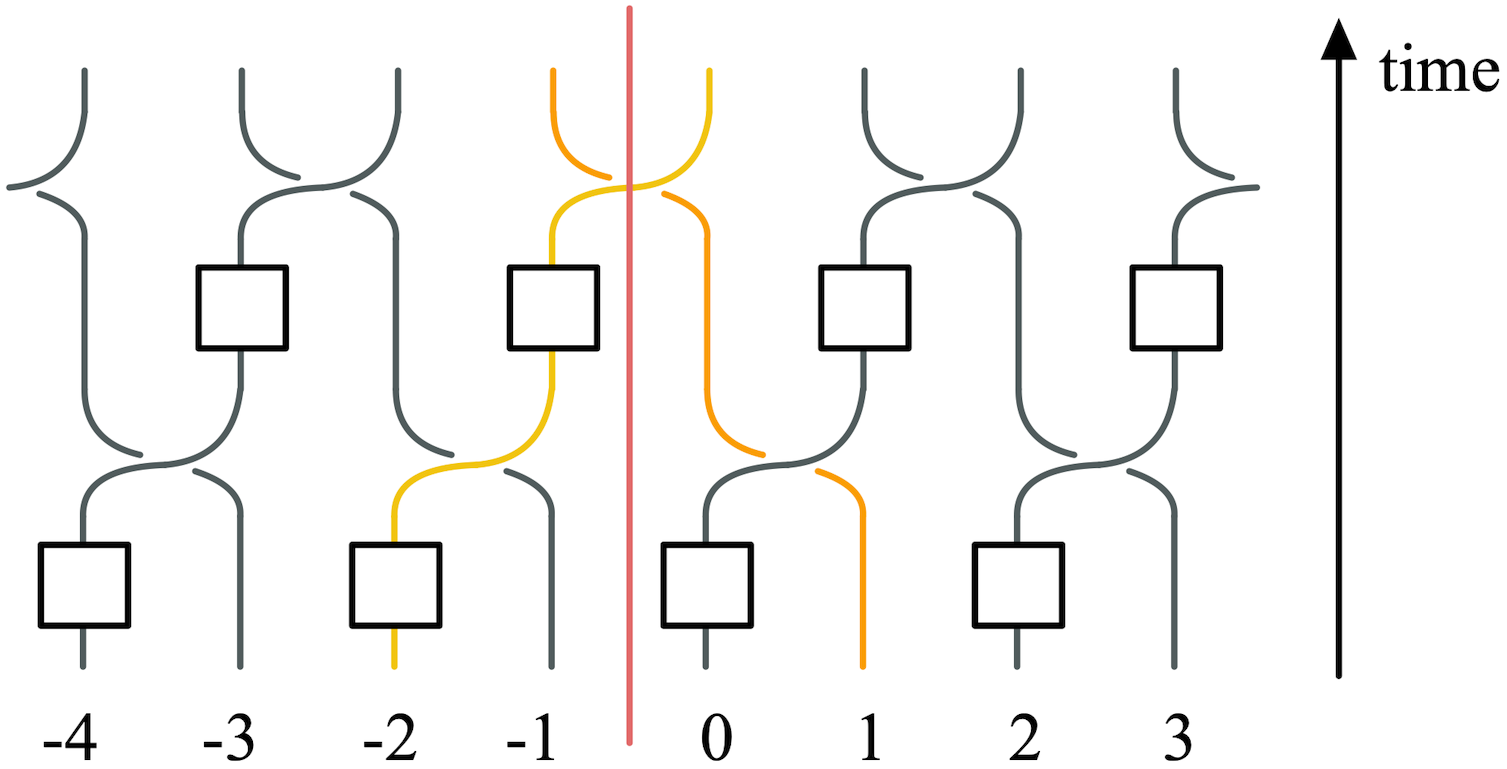}
    \caption{Illustration of one step of a local non-unitary QCA, the reset-swap map described in Sec. \ref{sec:resetswap}, consisting of local reset onto the state $\dyad{0}$ on every second site (indicated by boxes) followed by a pair-wise swap operation.  Since only the $\dyad{0}$ operator is propagated to the right while all single site operators are propagated to the left there is a net flow of information to the left across the boundary (pink vertical line) in distinction to local unitary QCA where there is none.}
    \label{fig:intro}
\efig

Referring to Fig.~\ref{fig:intro}, the pink vertical line in the center represents the boundary across which the information flow shall be measured.
The QCA acts on two qubits at a time, updating nearest neighbor pairs of lattice sites at locations $\{2j,2j+1\}_{j\in \mathbb{Z}}$, followed by the same update on pairs shifted by one lattice site. The composition of the two updates together constitutes a single time step of the QCA.
Each two-qubit update consists of three local operations: first, a quantum channel, indicated by a box, which resets the qubit on the left hand site of the pair to the state $\ket{0}$; second, an identity operation on the right cell, indicated by a straight line; and third, a swap operation, represented by crossed lines, which swaps the locations of the neighboring qubits.
One can see that only the yellow and orange colored worldlines of the initial operators at sites -2 and 1 cross the boundary after one QCA step.
Operators on the yellow path are reset by the map and thus only the operator $\{\dyad{0}\}$ is transported to the right,
while on the orange path all four orthonormal operators $\{\op{0},\op{0}{1},\op{1}{0},\op{1}\}$ are transported to the left.
Thus, in distinction to the unitary case in which no information flow is present, a net flow of quantum information occurs to the left for this non-unitary local QCA.

In the next chapter, we show how to define such a measure which captures the information flow in discrete, translationally-invariant systems independent of the type or dimension of the quantum state.

\section{Quantifying information flow}

In the following, the mathematical background of the MPO description of QCA is presented in Sec.~\ref{sec:MPOdescription}, before providing a short summary of the on MPUs based index theory for unitary QCAs in Sec.~\ref{sec:indextheory}.
Sec.~\ref{sec:current} outlines the definition of the information current for non-unitary QCA, whose properties are listed and discussed in the final Sec.~\ref{sec:properties}.

\subsection{MPO description of QCA} \label{sec:MPOdescription}

In this framework, a single time step of the QCA is modeled by an MPO in the most general form, see Fig.~\ref{fig:QCAupdate}(a), with the same local tensor $M$ of the MPO distributed equally across the lattice.
These tensors represent superoperators acting on vectorized density matrices in a doubled Hilbert space $\mathcal{H}\times\mathcal{H}^*$.

The MPO in Fig.~\ref{fig:QCAupdate}(a) represents the general form of the dynamical map. It can be represented by the circuit shown in Fig.~\ref{fig:QCAupdate}(b) when the QCA is exclusively defined by \textit{local} operation. This class of QCA will be referred to as ``local QCA'' throughout this work.
In the referred circuit, local operators $V$ (framed in yellow) act on pairs of neighboring sites, after which the set of operators $W$ (marked with red) update the next pairs of neighboring sites, shifted by one lattice site.
This is the simplest, and most commonly used partitioning scheme of QCA with a two-cell neighborhood. Note that generality is provided nonetheless, as the sites of QCA with larger interaction neighborhoods can be grouped together, such that it has the same structure as a QCA with a two-cell neighborhood (analogous to a coarse-graining process).
Further, the singular value decomposition (SVD) is applied by rewriting one of the operators that acts on two neighboring sites, e.g.~$V$, into a single index sum of tensor products of operators $B$ and $A$ which act on the associated left or right site, respectively, according to Fig.~\ref{fig:QCAupdate}(c).
Then the local tensors $M$ in Fig.~\ref{fig:QCAupdate}(a) can be defined according to Fig.~\ref{fig:QCAupdate}(d) --
as constituent four-index tensors, whose (vertical) physical indices have been grouped together.

\bfig
    \includegraphics[width=.97\columnwidth]{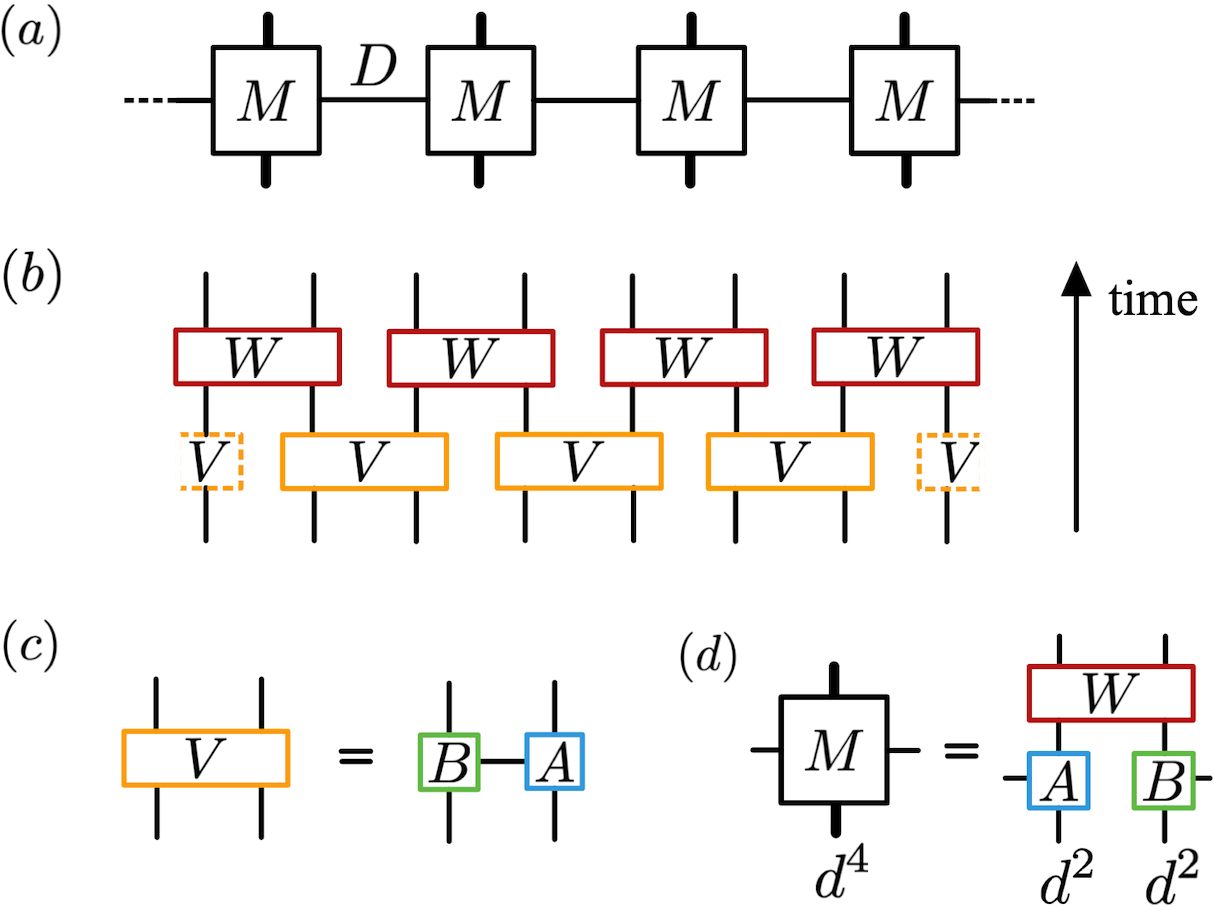}
    \caption{(a) One time step of a QCA, written as an MPO with virtual bond dimension $D$.
    It acts on a doubled Hilbert space $\mathcal{H}\times\mathcal{H}^*$, and is in (b), (c), and (d) assumed to be locally generated. Its action on two qudits is represented as a vectorization of operators with physical dimension $(d^2)^2=d^4$.
    (b) Partitioning scheme of the MPO in (a) for local QCA (excluding e.g.~the shift map).
    (c) Singular value decomposition of the tensor operator $V$. Note that the horizontal bond dimension $D$ is equal to the rank of the operator singular value decomposition of $V$.
    (d) Definition of the constituent local tensor $M$ of the MPO.}
    \label{fig:QCAupdate}
\efig

On the basis of the MPU description of QCA, an index theory has been formulated for unitary QCA; it is summarized below including a reformulation of its definition.

\subsection{Index theory for unitary QCA using MPUs} \label{sec:indextheory}

Following Refs.~\cite{PhysRevB.98.245122,Ignacio_Cirac_2017}, we define the matrices $M_L$ and $M_R$ with input and output Hilbert spaces as indicated in Fig.~\ref{fig:MLandMR}.
In \cite{Ignacio_Cirac_2017} it is shown that for unitary one-dimensional QCA, one can quantify the net flow of quantum information to the right via the so-called rank-ratio index \footnote{Throughout we take logarithms base 2.}:
\ba
    \rm ind
    &= \frac{1}{2}\qty(\log{\Rank(M_R)}-\log{\Rank(M_L)}) \nn\\
    &= \frac{1}{2}\log \l(\frac{\Rank(M_R)}{\Rank(M_L)}\r)
    .
\label{RankIndex}
\ea
In anticipation of our alternative measure for information flow below, we note that because $\Rank(A)=\Rank(A^\dagger A)$ for any complex matrix $A$, the index can also be written as
\ba
    {\rm ind}
    &=\frac{1}{2}(S_0(\sigma_R)-S_0(\sigma_L)),
\ea
where
\begin{equation}
\sigma_{\beta}=M^{\dagger}_{\beta}M_{\beta}/\Tr[M^{\dagger}_{\beta}M_{\beta}],\quad {\rm for \ } \beta\in\{L,R\},
\label{eq:sigmaop}
\end{equation}
are trace one, positive, Hermitian operators.
Here 
$S_0(\rho)={\rm\log Rank}(\rho)$, also known as the Hartley entropy,
is the $\alpha=0$ case of the R\'enyi-$\alpha$ entropy
\begin{equation}
S_{\alpha}(\rho)=\frac{1}{1-\alpha}\log \Tr[\rho^{\alpha}].
\end{equation}

\bfig
    \includegraphics[scale=.21]{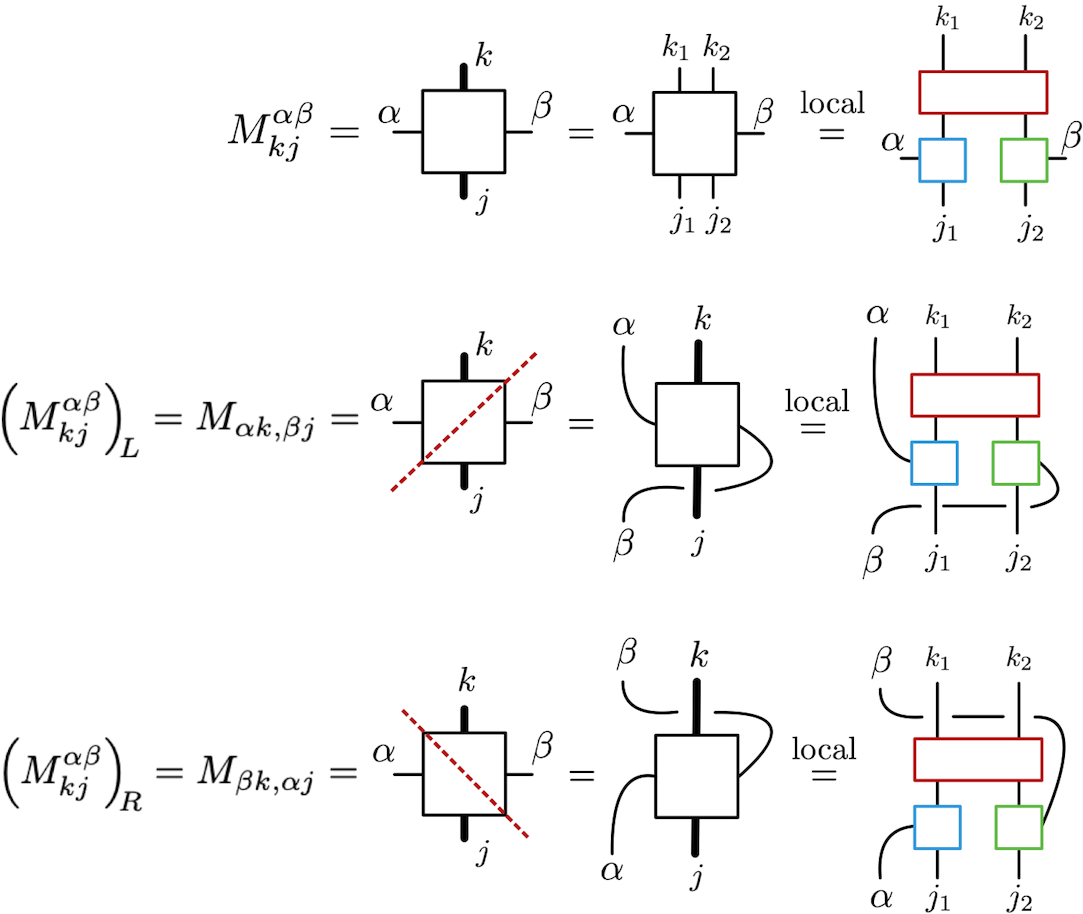}
    \caption{Diagrammatic description of the matrix components of $M_L$ and $M_R$ obtained by taking left and right partitionings of the local tensor $M$ of the MPO that describes the QCA. $\alpha$ and $\beta$ are \emph{virtual} indices, while $j=(j_1,j_2)$ and $k=(k_1,k_2)$ represent the composed \emph{physical} indices of the input or output state, respectively. The adjoint matrix components  $\Big(M_{kj}^{\alpha\beta}\Big)_{\!\!\:\!L,R}^{\!\dagger}$ are obtained by reflecting the diagrams about the horizontal axis and replacing the constituent tensors by their adjoints.
    Note the equations on the furthest right site are only true if the QCA is local.
    } 
    \label{fig:MLandMR}
\efig

The index has been shown to be a rigid quantity in the sense that all \textit{locally equivalent} unitary QCA, i.e.~those QCA that are obtainable from each other by a finite-depth sequence of local QCA updates, have the same index. In particular, for locally generated unitary QCA, $\ind =0$, while for non-locally generated unitary QCA the index is a positive rational. The latter include for example the shift operation, for which the index is roughly defined by the fraction of the number of shifts to the right divided by the number of shifts to the left.


\subsection{Information current in non-unitary QCA} \label{sec:current}

For non-unitary QCA, the index is no longer a rigid quantity as it does not remain invariant under local operations when coupling the system to the environment.
We seek a quantity which captures information flow in non-unitary QCA, but which is zero for local unitary QCA.

This quantity should be continuous with the parameter that describes the coupling to the environment since non-unitary dynamics can be continuously connected to unitary dynamics.
A natural quantity to consider, extending Eq.~\eqref{RankIndex}, is a continuous function on the singular values of $M_L$ and $M_R$.
Note the values and the total number of non-zero singular values of $M_L$ and $M_R$ can change.

To motivate such a quantity we give a couple observations.
First, the singular values of $M_L$ and $M_R$ are equal for local unitary QCA; see proof in App.~\ref{AppA}.  
Second, the trace of the first moments of $M_R^{\dagger}M_R$ and $M_L^{\dagger}M_L$ are equal for \emph{all} QCA; see Fig.~\ref{fig:firstmoment}.
\bfig
    \centering
    \includegraphics[scale=.15]{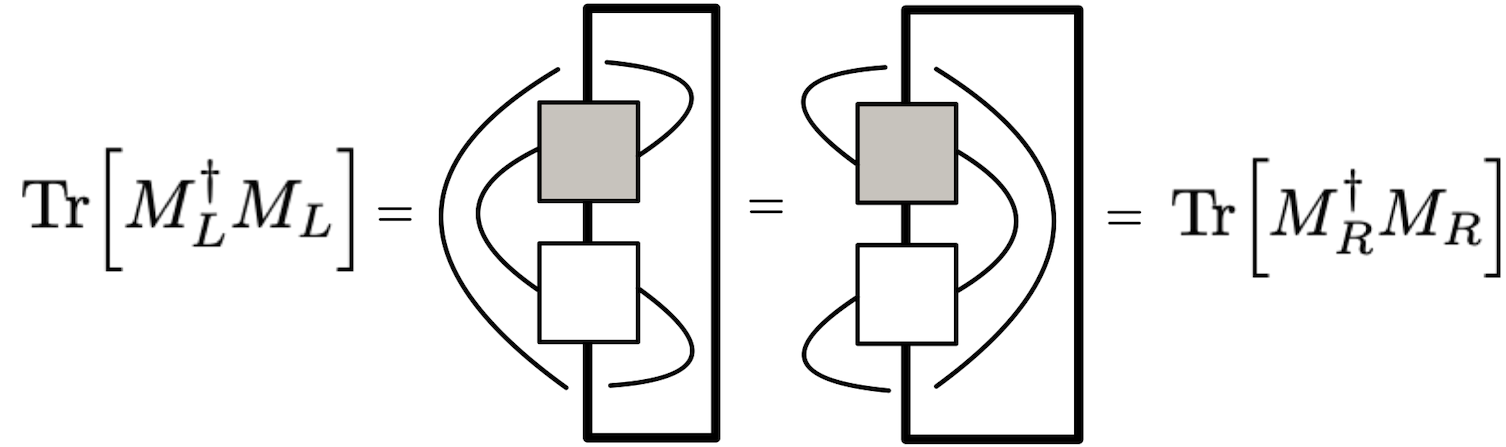}
    \caption{Tensor network representing the equality of the traces of the first moments of $M_L^{\dagger}M_L$ and $M_R^{\dagger}M_R$. The equation holds for all QCA, independent of it being unitary or non-unitary.}
    \label{fig:firstmoment}
\efig

The lowest moment of the squared eigenvalues that can distinguish unitary and non-unitary dynamics is the second. Hence, we propose a measure of the information current, namely the information flow per update time increment, based on the \Renyitwo entropy of the operators $\sigma_{\beta}$ defined in Eq.~\eqref{eq:sigmaop}: 
\ba
    I
    &=\frac{1}{2}(S_2(\sigma_R)-S_2(\sigma_L))\nn\\
    &=\frac{1}{2}\log\l(\frac{\Tr[\l(M_L^\dagger M_L\r)^{\!2}]}
    {\Tr[\l(M_R^\dagger M_{R}\r)^{\!2}]}\r).
    \label{eq:current}
\ea
Thus, the current $I$ differs from the index simply by taking difference of \Renyitwo entropies rather than the R\'enyi-0 entropies.
The tensor network description of the argument of the logarithm is shown in Fig.~\ref{fig:current}.

In App.~\ref{sec:CJS} it is derived that the current can also be reformulated in terms of the difference in \Renyitwo entropies of the inner product of the Choi-Jamiolkowski state (CJS) associated with $M_L^\dagger M_L$ and $M_R^\dagger M_R$, respectively.

In order to calculate the current we need to construct the matrices $M_L$ and $M_R$ from the QCA rule.
As the operator $V$ is local, we can write 
\ba
    V = \sum_{r,s=1}^{d^2} c_{r,s} \, \hat O_r\otimes \hat O_s,
    \label{eq:V_SVD}
\ea
where $\{\hat O_r\}_{r=1}^{d^2}$ is any orthonormal basis for operators on a qudit satisfying $\Tr[\hat O_r^{\dagger}\hat O_{r'}]=\delta_{r,r'}$. Note when acting on vectorized density matrices, each operator $\hat O_r$ acts on this doubled space as $\hat O_{r}\otimes \hat O_r^{\ast}$.
A singular value decomposition can be performed on the matrix of coefficients
\ba
    c_{r,s} = \sum_{k=1}^{\chi}Y_{r,k}D_{k,k}X^{\dagger}_{k,s},
\ea
where $Y$ and $X$ are unitary, $D$ is the diagonal matrix of singular values of $c$, and the rank of the decomposition is $\chi$ with $1\leq \chi\leq d^2$.

\bfig
   \centering
    \includegraphics[scale=0.27]{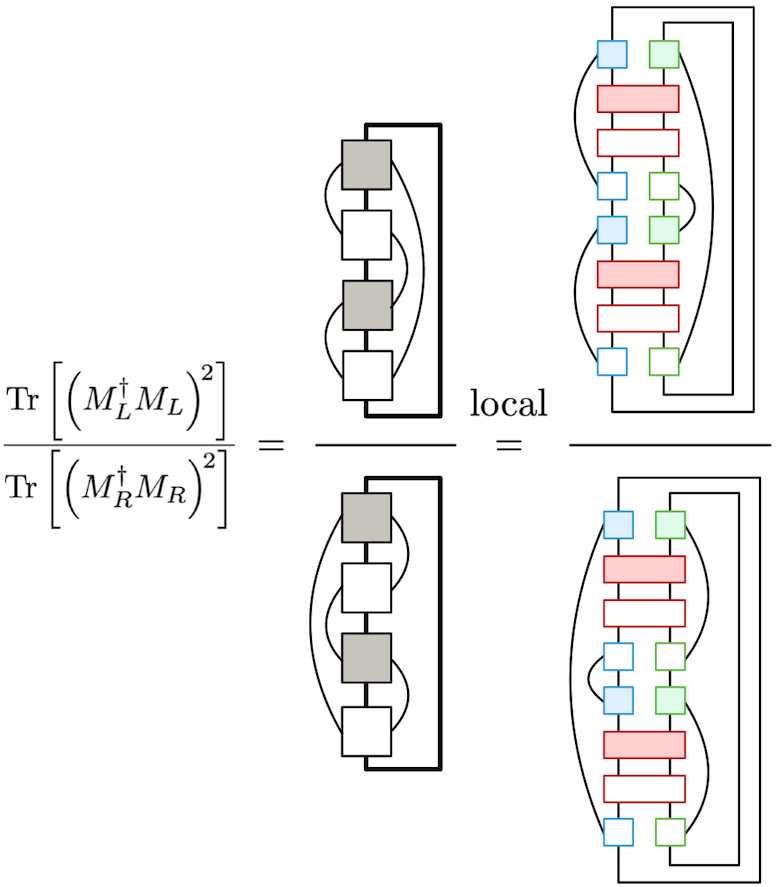}
    \caption{A measure of information current in QCA. The tensors are as defined in Fig.~\ref{fig:QCAupdate}, where the shaded tensors represent the associated adjoints. Note that the expression on the right-hand site is only valid for local QCA.}
    \label{fig:current}
\efig
Expanding we have
\ba
    V 
    &= \sum_{k=1}^{\chi} \sum_{r=1}^{d^2} Y_{r,k}\,\sqrt{D_{k,k}} \,\hat O_r\otimes \sum_{s=1}^{d^2} \sqrt{D_{k,k}} \,X^{\dagger}_{k,s}\,\hat O_s \nn\\
    &=:\sum_{k=1}^{\chi} B_k\otimes A_k,
\ea
where 
\ba
    A_k &= \sum_{s=1}^{d^2} \sqrt{D_{k,k}} \ X^{\dagger}_{k,s} \ \hat O_s, \ \
    B_k = \sum_{r=1}^{d^2} \sqrt{D_{k,k}} \ Y_{r,k} \ \hat O_r.
    \label{localtensors}
\ea
Note the equally weighted symmetric distribution of the singular values on the two local matrices $A$ and $B$, with both exhibiting a factor of $\sqrt{D_{k,k}}$.
This assures that the same magnitude of the current is observed after a parity operation on the QCA rule – only the sign of the current would be reversed.

Applying the singular value decomposition of $V$ according to Eqs.~\eqref{eq:V_SVD} to \eqref{localtensors}, the current can be for local QCA explicitly written in terms of the constituent tensors of $M_L$ and $M_R$ ($W$, $A$, and $B$, see Fig.~\ref{fig:QCAupdate}(b)):
\begin{widetext}
\ba
    I
    &=\frac{1}{2}\log\l(\frac{\Tr[
    \sum_{a,b,c,d=1}^\chi 
    \l(A_a^\dagger\otimes B_b^\dagger\r)
    W^\dagger W
    \l(A_a\otimes B_d\r)
    \l(A_c^\dagger\otimes B_d^\dagger\r)
    W^\dagger W
    \l(A_c\otimes B_b\r)
    ]}{\Tr[
    \sum_{a',b',c',d'=1}^\chi 
    \l(A_{a'}^\dagger\otimes B_{b'}^\dagger\r)
    W^\dagger W
    \l(A_{c'}\otimes B_{b'}\r)
    \l(A_{c'}^\dagger\otimes B_{d'}^\dagger\r)
    W^\dagger W
    \l(A_{a'}\otimes B_{d'}\r)
    ]}\r).
    \label{eq:I}
\ea
\end{widetext}

\subsection{Properties of the information current} \label{sec:properties}

The main properties of the current are summarized in the points below. They are to be compared with the index theorem in \cite{index_werner}, which has been proposed as an equivalence class for unitary QCA. The associated quantity is here referred to as the ``GNVW index''.
Note that this index theorem has already been extended and generalized in 
a corresponding MPU description \cite{Ignacio_Cirac_2017}, 
a fermionic version \cite{gong2020topological}, 
and
a generalization to approximately locality-preserving QCA \cite{ranard2020converse}. 

%

\subsubsection{\it{I} is locally computable.} \label{sec:local}
As any QCA can be fully characterized by the locality-preserving operators $M$ of an MPO,
and $I$ is a function of $M$, it is locally computable.
The same property has been shown for the GNVW index \cite{index_werner}.

\subsubsection{\it{I} is vanishing for unitary finite-depth circuits.} \label{sec:unitary}
No information flow is present in the case of unitary finite-depth circuits, which do not include shift operations nor interactions of the system with the environment. Mathematically, this is shown for local unitary QCA in Fig~\ref{fig:unitaryproof}, where $I =\frac{1}{2}\log(1)=0$ following Eq.~\eqref{eq:current}.
The pictured equation is obtained by setting $W^\dagger W = \mathds{1} \otimes \mathds{1} = V^\dagger V$ and using $\sum_{k=1}^{\chi} A_k A^{\dagger}_k=\sum_{k=1}^{\chi} B_k B^{\dagger}_k=\sum_{k=1}^{\chi} A^{\dagger}_k A_k=\sum_{k=1}^{\chi} B^{\dagger}_k B_k=c \mathds{1}$; see derivation in App.~\ref{AppA} up to Eq.~\eqref{eq:unitarycondition}.
In addition, note that the singular values of $M_L$ and $M_R$ are in this case equal as shown further in App.~\ref{AppA}.
    \bfig
        \centering
        \includegraphics[scale=0.28]{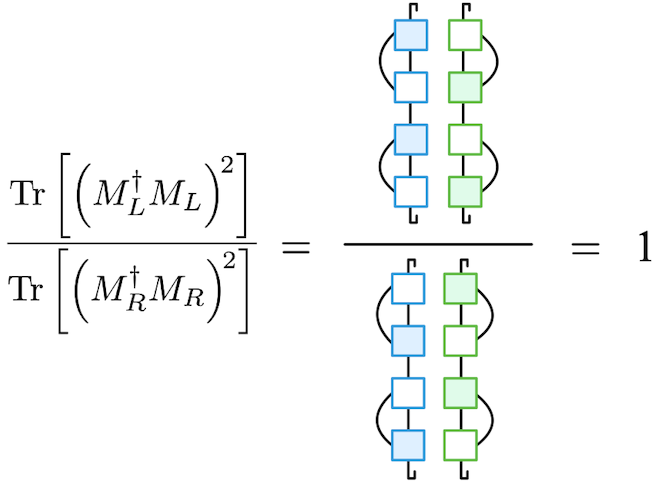}
        \caption{On property \ref{sec:unitary}: Argument of $I$ as in Fig.~\ref{fig:current}, but under the assumption of local unitary QCA such that $W^\dagger W=\mathds{1}$. For clarity the traces are here indicated by ``hooks" on the top and on the bottom of the four tensor chains, and the cyclic property of the trace has been used for the $B$'s in the numerator and the $A$'s in the denominator.
       }
        \label{fig:unitaryproof}
    \efig
The property has also been proven for the GNVW index \cite{index_werner}.

\subsubsection{\it{I} is not invariant under blocking.} \label{sec:blocking}
The blocking procedure describes the regrouping of all physical sites of a QCA, 
where two or more neighboring sites are grouped together to define a supercell.
    One could think of this as a coarse-graining procedure, and has been shown to not change the dynamics of a QCA.
    Here, the blocking is described by taking the tensor product of two or more local tensors of the associated MPO: $M\rarrow M^{\otimes^n}$, where $n\in\mathbb{N}_{\geq2}$.
    The corresponding tensor network description is presented in Fig.~\ref{fig:currentblocking}. 
    \bfig
    \centering
    \includegraphics[width=\columnwidth]{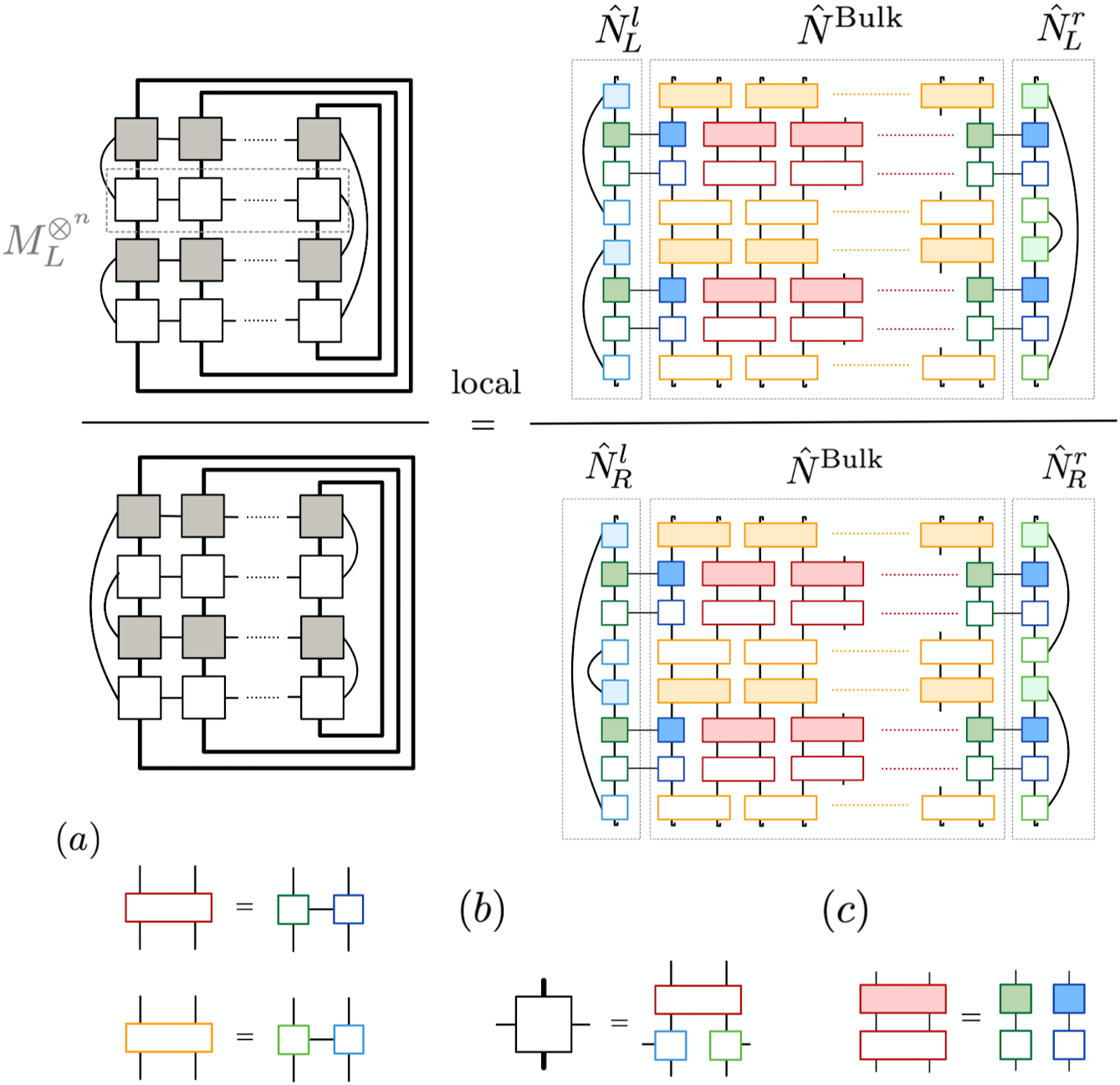}
    \caption{
    On property \ref{sec:blocking}:
    Tensor network description of the blocked version of the information measure shown in Fig.~\ref{fig:current}.
  The expression on the left, $\frac{\tx{Tr}\!\left[\qty(\left(M^{\otimes^n}\right)_{\!_L}^\dagger \left(M^{\otimes^n}\right)_{\!_L})^2\right]}
    {\tx{Tr}\!\left[\qty(\left(M^{\otimes^n}\right)_{\!_R}^\dagger \left(M^{\otimes^n}\right)_{\!_R})^2\right]}$, holds for all QCA, whereas the one on the right, $\frac{\mel**{\hat N_{\!L}^l}{\hat N^\tx{Bulk}}{\hat N_{\!L}^r}}{
    \mel**{\hat N_{\!R}^l}{\hat N^\tx{Bulk}}{\hat N_{\!R}^r}}$, can only be applied to local QCA.
     $\hat N_{\!L}^l$ and $\hat N_{\!R}^l$ ($\hat N_{\!L}^r$ and $\hat N_{\!R}^r$) label the grouped tensors in the numerator and denominator, respectively – they act on the first (last) sites, including their associate traces and sums over their virtual indices.
     The composition of all other tensors acting on the physical sites in the middle are labeled by $\hat N^\tx{Bulk}$. 
     Note that this tensor is the same for the numerator and the denominator,
     and cancel each other out if condition (c) is fulfilled: $W^\dagger W$ is factorizable (or equivalently if $\mathrm{Rank}(W^\dagger W)=1$), such that there is no virtual bond connection between $\hat N^\tx{Bulk}$ and the boundary tensors.
     In this case the tensor diagram equals the one in Fig.~\ref{fig:current}, proving that $I$ is invariant under blocking if $W^\dagger W$ is factorizable.
     (a) Definition used to illustrate the singular value decomposition of the local gates $W$ (top) and $V$ (bottom).
     (b) Definition of the composed local tensors $M$.
     }
    \label{fig:currentblocking}
\efig

    Fig.~\ref{fig:currentblocking} shows that $I$ does not change under blocking if $W^\dagger W$ is factorizable, see condition in subfigure \ref{fig:currentblocking}(c),
    while $I$ is not necessarily invariant under blocking
    if $W^\dagger W$ is \textit{non-}factorizable.
    The latter does not meet expectations, as the dynamics of a (translation-invariant) QCA have been shown to be invariant under blocking.
    %
    However, it is possible to specify certain conditions under which the current does stay invariant under the blocking procedure --- say, if the tensor $V$ is unitary, $I$ would remain invariant under increasing the number of blocked sites from four to more even numbered sites, i.e.~six, eight, or ten.
    Further, it is to highlight that $I$ does not change by blocking for all unitary QCA, which can be described by the set of all finite-depth circuits and shift maps: The former is in Sec.~\ref{sec:unitary} defined by $W^\dagger W=\mathds{1}\otimes\mathds{1}$, which is factorizable, and the latter is shown to be invariant under blocking in Sec.~\ref{sec:shift}.
    This is in alignment with the GNVW index introduced in Ref.~\cite{index_werner} which has been shown to be independent of how we regroup or block sites of the unitary QCA.
    
    The invariance under blocking for particular QCA is presented in the next section \ref{sec:examples}, where both, QCA for which $W^\dagger W$ is factorizable (the shift and the reset-swap QCA), and QCA with non-factorizable $W^\dagger W$ (assuming the directed amplitude damping channel) are discussed.


\subsubsection{\it{I} is not additive under composition.} \label{sec:composition}
Similar to the blocking procedure, where physical cells are grouped together to supercells, one can also compose two or more QCA updates: $M\rarrow \qty(M^{\otimes^n})^n$, where the $M^{\otimes^n}$ acts on $n$ physical sites $n$ times.
To compute the current for a composition of $n$ time steps of the QCA, one has to take $n$ lattice sites into account. This can be understood by the causal cone structure of QCA that defines a finite bound of information flow in the system.
One would loose information if one were to consider a composition of QCA without also the blocking of sites.
Considering the composition of $n=2$ QCA updates, the dynamics are captured in the corresponding tensor network description in Fig.~\ref{fig:currentcomposition}.
\begin{figure}
    \includegraphics[scale=0.29]{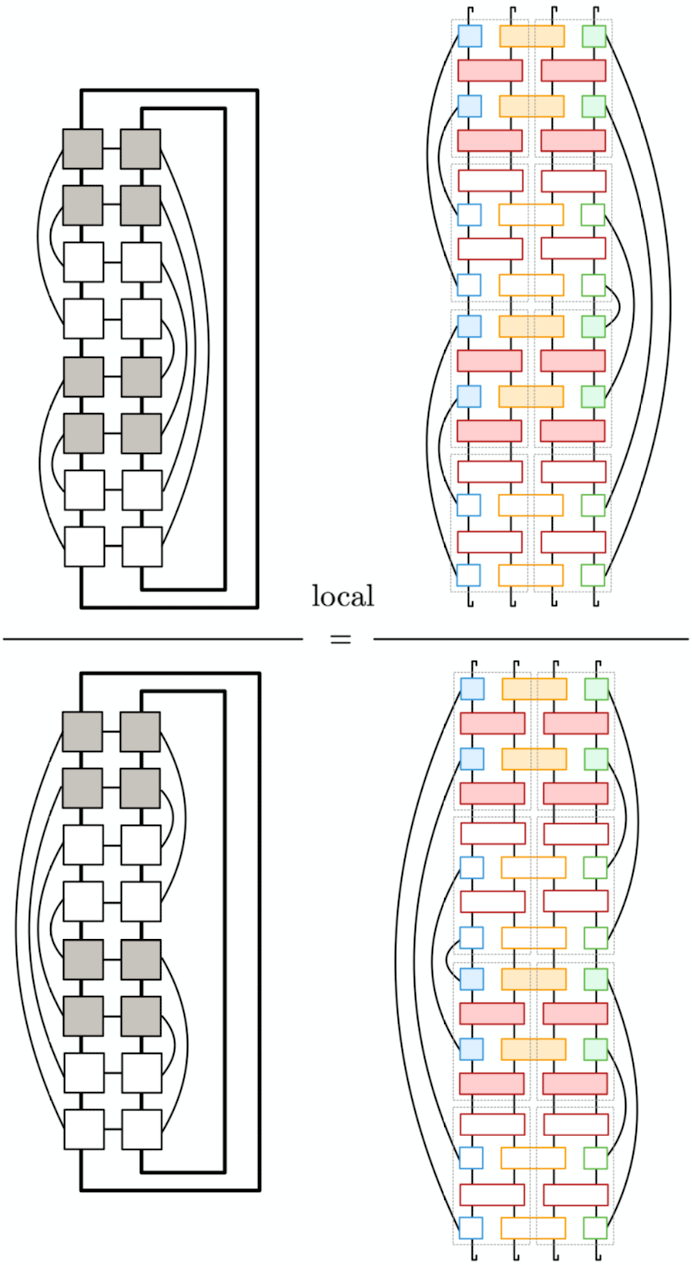}
    \caption{
    On property \ref{sec:composition}:
    Argument of the current for the composition of two QCA updates including the blocking of two sites:
    $\frac{\tx{Tr}\!\qty[\qty(\qty(M^2\otimes M^2)_{\!_L}^\dagger \qty(M^2\otimes M^2)_{\!_L})^2]}
    {\tx{Tr}\!\qty[\qty(\qty(M^2\otimes M^2)_{\!_R}^\dagger \qty(M^2\otimes M^2)_{\!_R})^2]}$.
    The light gray dashed boxes in the diagram on the right-hand site indicate the constituent tensors of $M^2$ (or $\qty(M^2)^\dagger$ if the tensors are shaded).
    }
    \label{fig:currentcomposition}
\efig
    
The scaling of the current with the number of compositions depends on the class of QCA.
For unitary finite-depth circuits, for example, the information current remains vanishing independent of the number of compositions.
For the shift map, or the (classical) full-reset-SWAP QCA with $p=1$, however, it is shown that $I$ is additive with the number of composed QCA updates; see Sec.~\ref{sec:examples}.
This additivity under composition has also been observed for the GNVW index for all unitary QCA
\footnote{Note that in the original paper \cite{index_werner} it was stated that index was multiplicative (not additive) under tensoring, which has been later corrected.}.
Other instances of QCA feature either super- and subadditivity, as discussed in Sec.~\ref{sec:examples},
which show that there does not exist a general scaling of the current under composition.
    
\subsubsection{\it{I} is vanishing if \it{W} and \it{V} are swap symmetric.} \label{sec:swapsymmetric}
Swap-symmetric QCA are described by maps whose action is the same on both, the left and the right site of the local neighborhood they act on. This means the system is spatially symmetric and it is impossible to define a certain direction of the information flow. 
Underpinning the intuition, the information current is shown to be vanishing for swap-symmetric QCA by the pictorial proof in Fig.~\ref{fig:swapsymmetricproof}.

\begin{figure}
    \centering
    \includegraphics[width=.77\columnwidth]{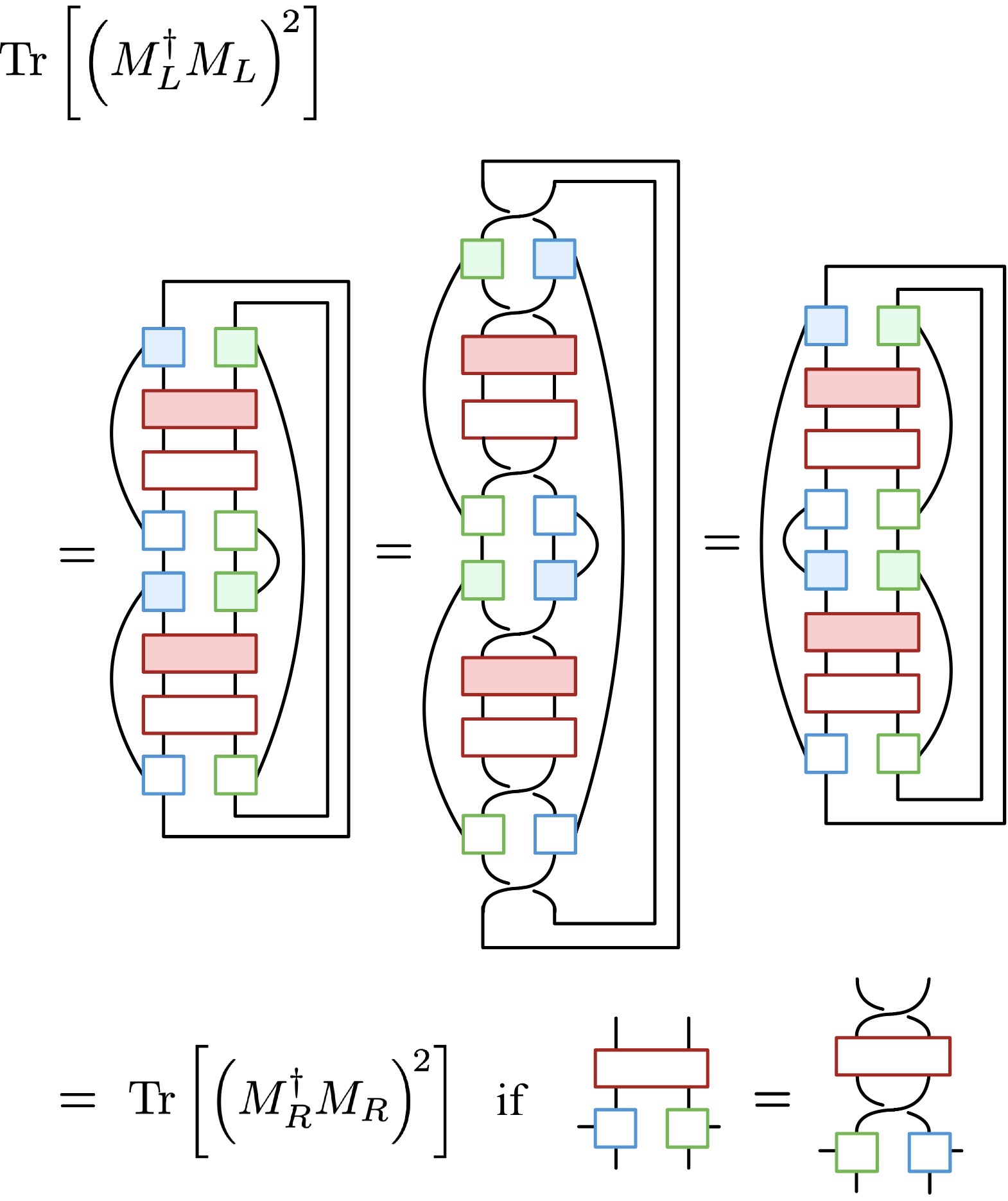}
    \caption{On property \ref{sec:swapsymmetric}: Proof of $I=0$ for swap-symmetric quantum channels, see condition on the bottom right.
    In the first step, the definition of $\Tr[\l(M_L^\dagger M_L\r)^{\!2}]$ on the top right in Fig.~\ref{fig:current} is applied.
    In the second step, the swap-symmetry condition is used.
    In the third step, the swap operators are applied onto the (blue and green framed) tensors $A$ and $B$, which swap the physical sites they act on.
    In the last step, the definition of $\Tr[\l(M_R^\dagger M_R\r)^{\!2}]$ on the bottom right in Fig.~\ref{fig:current} is rediscovered.
    The current is thus vanishing according to its definition in Eq.~\eqref{eq:current}, $I =\frac{1}{2}\log(1)=0$.
    }
    \label{fig:swapsymmetricproof}
\end{figure}


\section{Examples}\label{sec:examples}

In the following, the information flow in six different types of QCA is outlined.
First, the shift map in Sec.~\ref{sec:shift} serves as the standard example of non-local unitary QCA exhibiting non-zero information flow.
After that, two examples of non-unitary QCA are shown for which $W^\dagger W$ is factorizable: the reset-swap map as well as the dephase-swap map in Sec.'s~\ref{sec:resetswap} and \ref{sec:dephaseswap}.
Next, the directed amplitude damping channel in Sec.~\ref{sec:amplitudedamping} shows how the current behaves if $W^\dagger W$ is non-factorizable.
An instance of the property \ref{sec:swapsymmetric} is provided by the asymmetric swap map in Sec.~\ref{sec:asymmetricswap}, and the in  integrable models that satisfy the Yang-Baxter equations are studied in the last subsection \ref{sec:integrable}.

All investigated maps are categorized in Tab.~\ref{tab:examples} according to whether they are local or non-local, unitary or non-unitary, and whether they exhibit a zero or non-zero current. The shift map serves as a unique example of a unitary map with non-zero current due to its non-locality property. All other listed maps are non-unitary and local.
\begin{table}[h]
\begingroup
\setlength{\tabcolsep}{10pt}
\renewcommand{\arraystretch}{1.5}
\begin{center}
\begin{tabular}{ c| c c c c c c }
       & A & B & C & D & E & F \\ \hline
local  & \xmark & \cmark & \cmark & \cmark & \cmark & \cmark\\ 
NU & \xmark & \cmark & \cmark & \cmark & \cmark & \cmark\\
$|I|>0$   & \cmark & \cmark & \cmark & \cmark & \xmark & cf. Tab.~\ref{tab:integrable}
\end{tabular}
\end{center}
\caption{Overview of the properties of the presented exemplary maps: locality-preservation, non-unitarity, and the presence of an information current $I$. Note that example F comprises several (integrable) local and non-unitary maps which are specified in more detail in Tab.~\ref{tab:integrable}.}
\label{tab:examples}
\endgroup
\end{table}

Specifying the quantum channels in \ref{sec:resetswap} to \ref{sec:integrable}, only circuits as illustrated in Fig.~\ref{fig:QCAupdate}(b) are considered, where, for simplicity, the local tensors acting on the different neighborhoods at consecutive time steps are set to be the same, i.e.~$W=V$.
They are formally defined by superoperators which act on vectorized states $|\rho\rangle$, and exhibit an ordering of spaces $\mathcal{H}_1\times\mathcal{H}^{\ast}_1\times \mathcal{H}_2\times\mathcal{H}^{\ast}_2$.
For arranging the correct ordering of the operator bases of the two sites, the instances of the maps $W$ given in the examples below are as $d^4\times d^4$ matrices and transformed as follows:
\ba
    W
    \rarrow
    (\mathds{1}\otimes\hat\Sigma\otimes \mathds{1}) \
    W \
    (\mathds{1}\otimes\hat\Sigma\otimes \mathds{1}),
    \label{eq:Wbasischange}
\ea
where
\ba
    \hat\Sigma=\sum_{j,k=0,1} \dyad{j}{k}\otimes\dyad{k}{j}
\ea
is the swap operator which acts onto the vectorized Hilbert space.

\subsection{Shift map (non-local unitary QCA)} \label{sec:shift}

The shift map is defined by a non-local unitary QCA rule that shifts the algebra uniformly to the right by one lattice site. Using the MPO description, it is defined by the local tensors of the MPO shown in Fig.~\ref{fig:shift}, where the physical and virtual indices are regrouped according to left/right partitioning scheme presented in Fig.~\ref{fig:MLandMR}.

\bfig
    \includegraphics[width=\columnwidth]{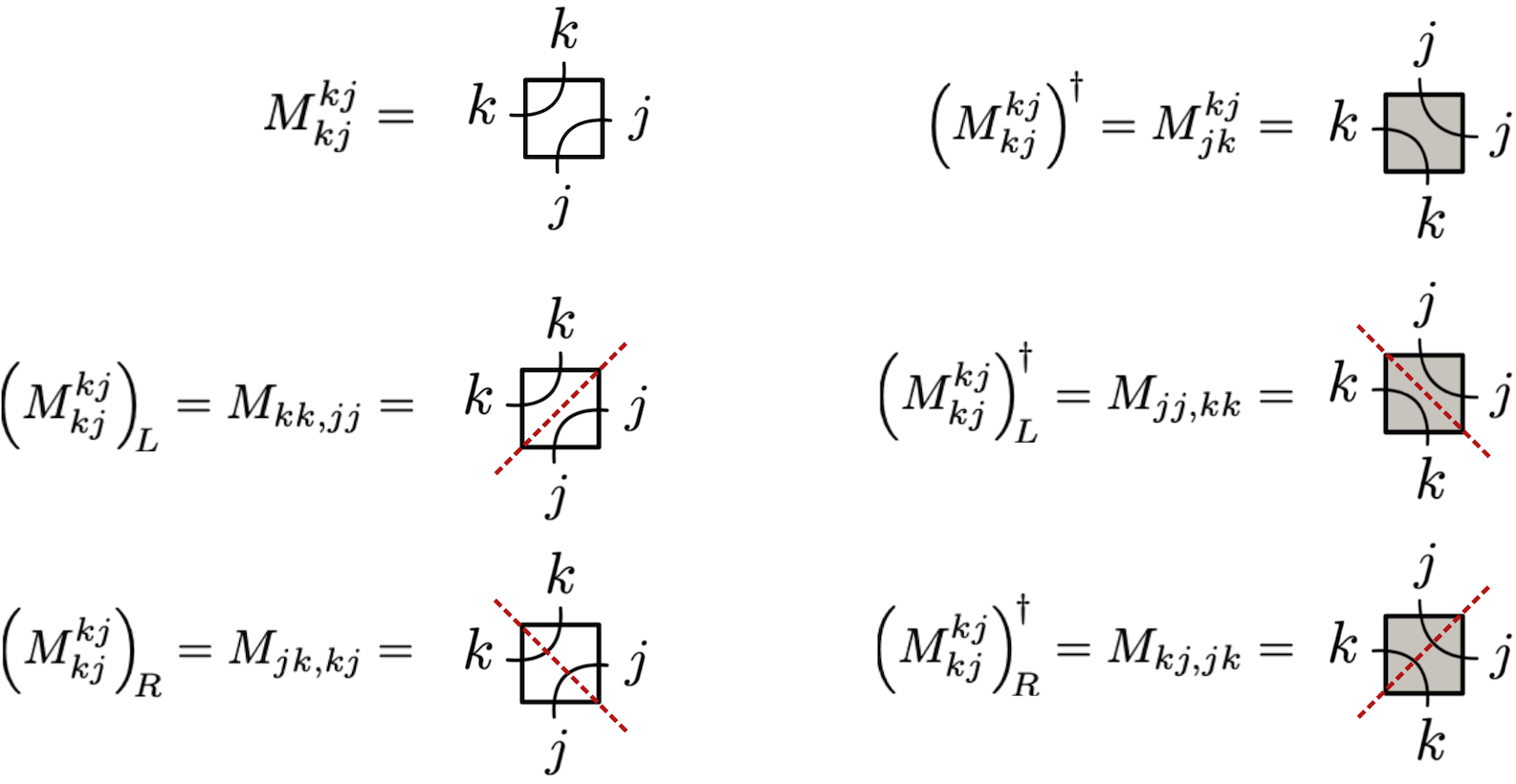}
    \caption{The non-local shift-right QCA. Left: definitions of the relevant tensor components of $M_{L,R}$.
    Right: respective adjoints of the same tensor components.
    Superscripts define the virtual bond dimensions while subscripts represent physical bond dimensions.}
    \label{fig:shift}
\efig

As proved diagrammatically in Fig.~\ref{fig:shiftcurrent}(a), the current for this map is $I=\log d^2$, which is the same as the index, $\tx{ind}=\log d^2$, since on the vectorized space $\Rank(M_R)=d^2 \cdot d^2=d^4$ while $\Rank(M_L)=1$
\footnote{Note that the index differs from the originally defined index by a factor of two as we are working in the doubled Hilbert space.}.
Under coarse-graining, where two lattice sites become one, the current does not change, see right-hand site of Fig.~\ref{fig:shiftcurrent}(a),
whereas it is shown to be additive under an additional composition of two QCA updates, see 
Fig.~\ref{fig:shiftcurrent}(b).

\bfig
    \includegraphics[width=\columnwidth]{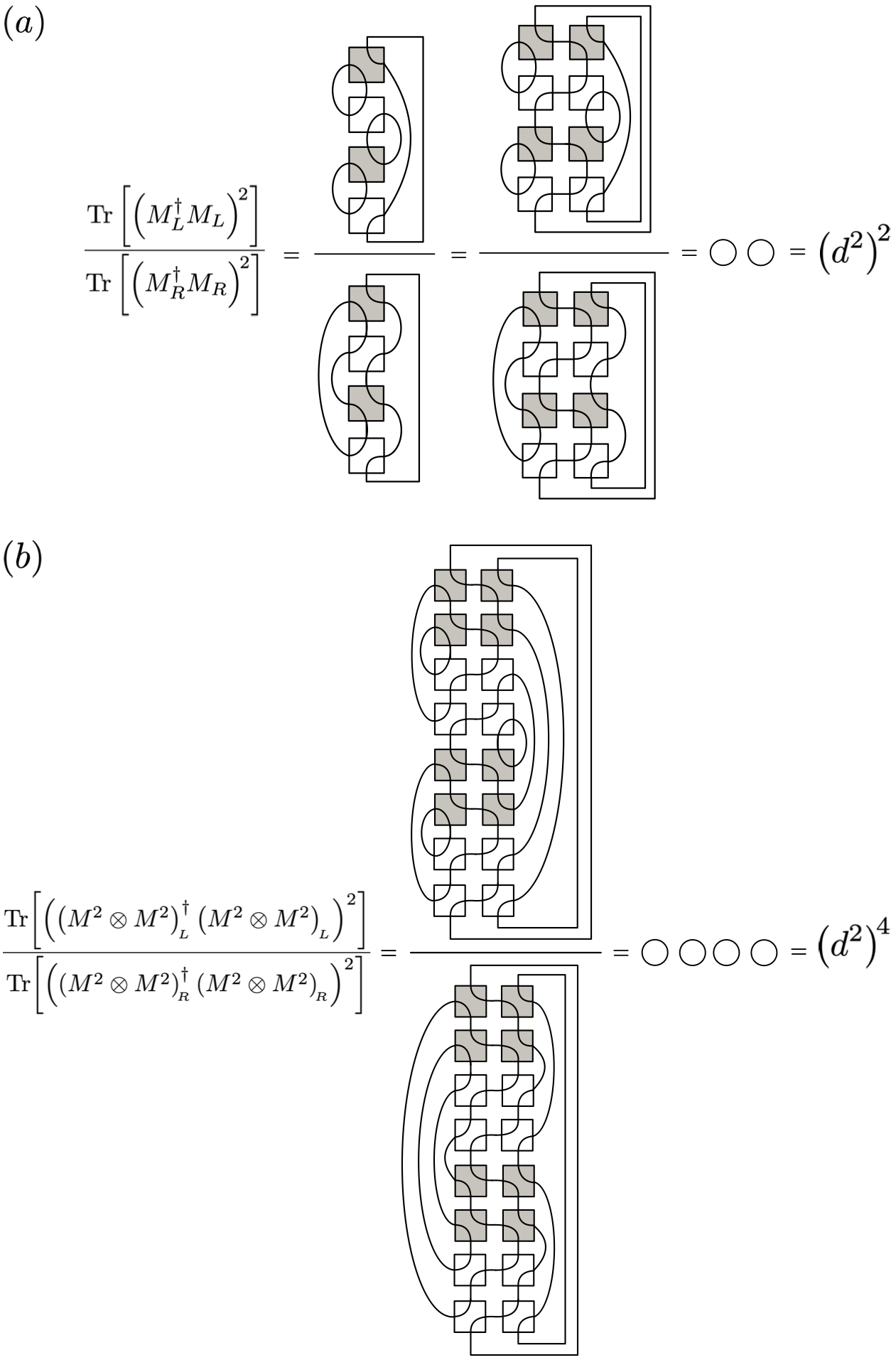}
    \caption{
    (a) Derivation of the current for the shift-right QCA: $I=\frac{1}{2}\log(d^2)^2=\log(d^2)$ remains invariant under course-graining (blocking), where two lattice sites become one.
    Note we work in a doubled space where each line represents the space $\mathcal{H}\times \mathcal{H}^{\ast}$, hence closed loops represent a trace which evaluates to $d^2$. Note that all the diagrams reduce to a number of closed loops.
    (b) Proof that under composition of $n=2$ shifts, including the blocking of two sites, the current increases by a factor of two: $I=\frac{1}{2}\log(d^2)^{\!4}=2\log(d^2)$.}
    \label{fig:shiftcurrent}
\efig

\subsection{Reset-swap map} \label{sec:resetswap}

The reset-swap map has been constructed to serve as a simple class of QCA which exhibit information flow due to its interaction with the environment, or spatially asymmetric loss of information.
Formally, it is a non-unitary quantum channel which acts on two sites: it (partially) resets either the left or the right cell, and swaps the two cells.
The corresponding tensor network description for the reset-left-swap map is illustrated in Fig.~\ref{fig:resetswap}, where, as discussed in the introduction around Fig.~\ref{fig:intro}, one could already suspect that there is a directional information flow present in the system by considering the ``flow'' of the basis operators of the algebras which describe the input states at the individual lattice sites --- one can see that there are more operators ``moving'' to one site than to the other.

The constituent tensors (or transfer matrices) which determine the local tensors $W$ of the associated MPO are defined in the following.
The reset operation is given by the two Kraus operators
\ba
K_0 =\m(1&0\\0&\sqrt{1-p}),\
K_1 =\m(0&\sqrt p\\0&0),
\label{eq:resetswapkraus}
\ea
which reset an input state to the $\dyad{0}$ state with probability $p\in[0,1]$.
In order to describe the corresponding action of the map onto operator algebras, the vectorization formalism in \cite{gilchrist2011vectorization} is used.
In this formalism, the reset gate is defined by 
\footnote{The vectorization follows from Eq.~(70) in \cite{gilchrist2011vectorization}. The conjugation of the tensors on the right-hand side of the tensor product is omitted as $\{K_\mu\}_{\mu=1,2}$ and $\hat\Sigma$ are real, $\hat\Sigma\in\mathbb R^{(2\times2)}$.}
\ba
    R = \sum_{\mu=0,1}K_\mu\otimes K^*_\mu,
\ea
while the swap superoperator is given by
\ba
    \tx{SWAP} = \hat\Sigma\otimes\hat\Sigma^*.
\ea
Then,
\ba
    W_{1,2} = \tx{SWAP}_{1,2} \, R_1
    \label{eq:W12}
\ea
represents the total reset-left-swap gate, where the subscripts define the physical sites on which the associated superoperators act on.
In the tensor network description used, this is the (red framed) tensor in Fig.~\ref{fig:QCAupdate}(d).
The single-site tensors $A$ and $B$ (framed in blue and green)
are obtained by applying the SVD onto the swap superoperator as follows:
\ba
    \tx{SWAP}_{1,2}
    =  \frac{1}{4}
    \sum_{a,b=0}^3 (\sigma^a \otimes \sigma^b)_1
    \otimes
    (\sigma^a \otimes \sigma^b)_2,
\ea
where $\{\sigma^a\}_{a=0}^3 = \{I,X,Y,Z\}$ is the set of Pauli operators.
Then one can write
\ba
    \tilde A_1^{(a,b)} = \frac{1}{2}  (\sigma^a\otimes\sigma^b)_1, \ \
    \tilde B_2^{(c,d)} = \frac{1}{2}  (\sigma^c\otimes\sigma^d)_2 \,R_2,
    \label{eq:A1_B2}
\ea
where the superscripts $(a,b)$ and $(c,d)$ are the virtual indices of $\tilde A_1$ and $\tilde B_2$, respectively, which arise from the SVD described above. (The operators are distinguished from above tensors $A$ and $B$ in Eq.~\eqref{localtensors} by a normalization factor that would account for a symmetric distribution of the singular values on the two operators --- the factor is dismissed here because it does not further change the result of the presented analytical derivation.)
In total, substituting Eqs.~\eqref{eq:W12} and \eqref{eq:A1_B2} in the definition of $M$, one obtains for the local MPO tensors:
\bs
\ba
    M^{(a,b,c,d)}_{1,2} 
    &= W_{1,2} \,\qty(\tilde A_1^{(a,b)} \otimes \tilde B_2^{(c,d)}) \nn\\
    &= \frac{1}{4} \text{SWAP}_{1,2} \, R_1 \, (\sigma^a\otimes\sigma^b)_1\otimes(\sigma^c\otimes\sigma^d)_2 \,R_2, \\
    \qty[M^{(a,b,c,d)}]^\dagger &= \frac{1}{4} R_2^\dagger \, (\sigma^a\otimes\sigma^b)_1\otimes(\sigma^c\otimes\sigma^d)_2 \, R_1^\dagger \, \text{SWAP}_{1,2}.
\ea
\label{eq:resetswapM}
\es

Using these definitions, the derivation of the current is presented in two different ways: first, diagrammatically using the tensor network description in Fig.~\ref{fig:resetswapcurrent}, and second, algebraically as a function of $R$ and $\{\sigma^a\}_{a=0}^{3}$ in App.~\ref{AppResetswap}.

\begin{figure}
    \includegraphics[width=\columnwidth]{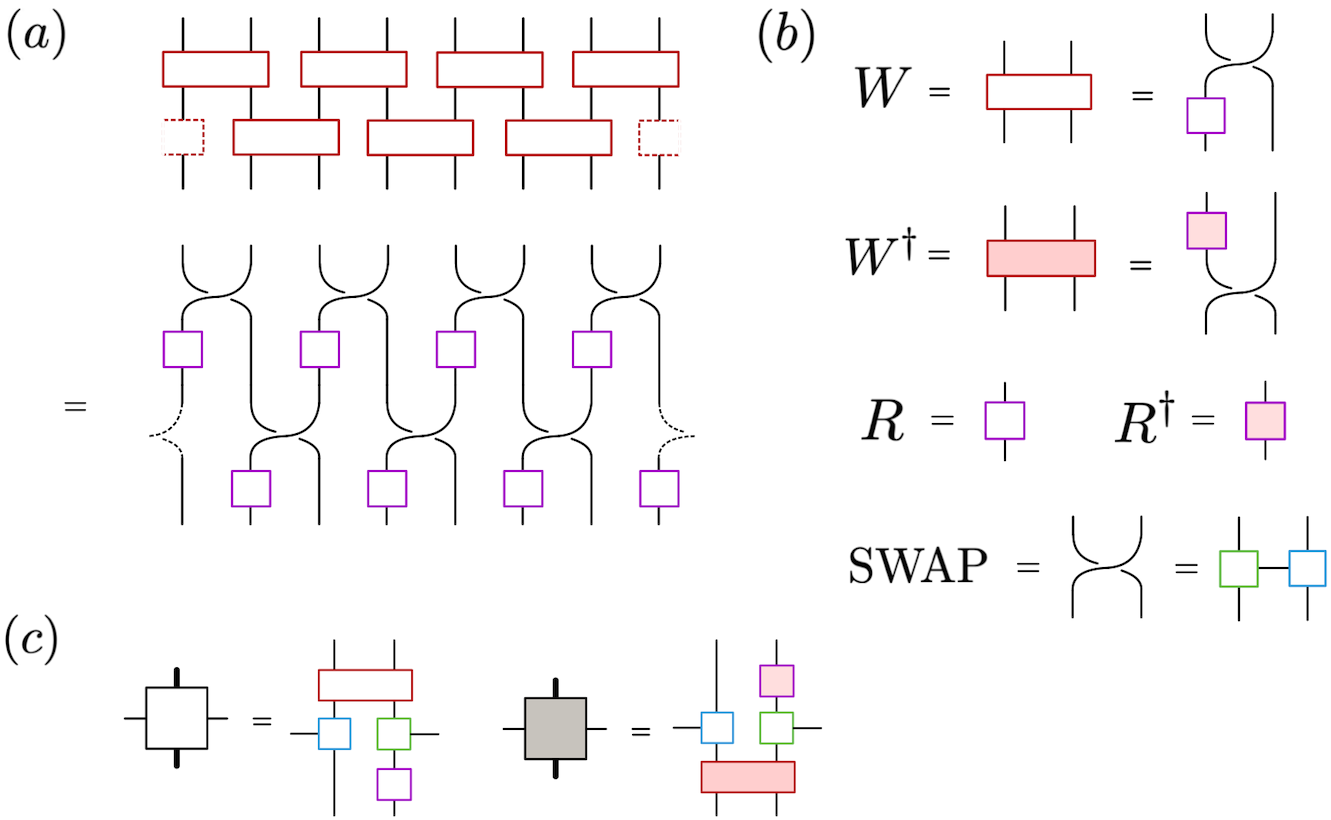}
    \caption{Illustration of the reset-swap QCA.
    (a) Circuit representation of one time step of the QCA.
    (b) Definition of the constituent local tensors 
    $W,W^\dagger,R,R^\dagger$, and $\text{SWAP}=\sum_{k=0}^{15}A_k\otimes B_k$.
    (c) Definition of the total local tensors $M$ and $M^\dagger$.
    Note that in the representation of $M^\dagger$ the single-site tensors $A$ and $B$ are not shaded as the Pauli operators that define the swap operation are self-adjoint; $\sigma^a=(\sigma^a)^\dagger \ \forall a\in[0,3]$.
    }
    \label{fig:resetswap}
\efig

\bfig
    \includegraphics[scale=0.3]{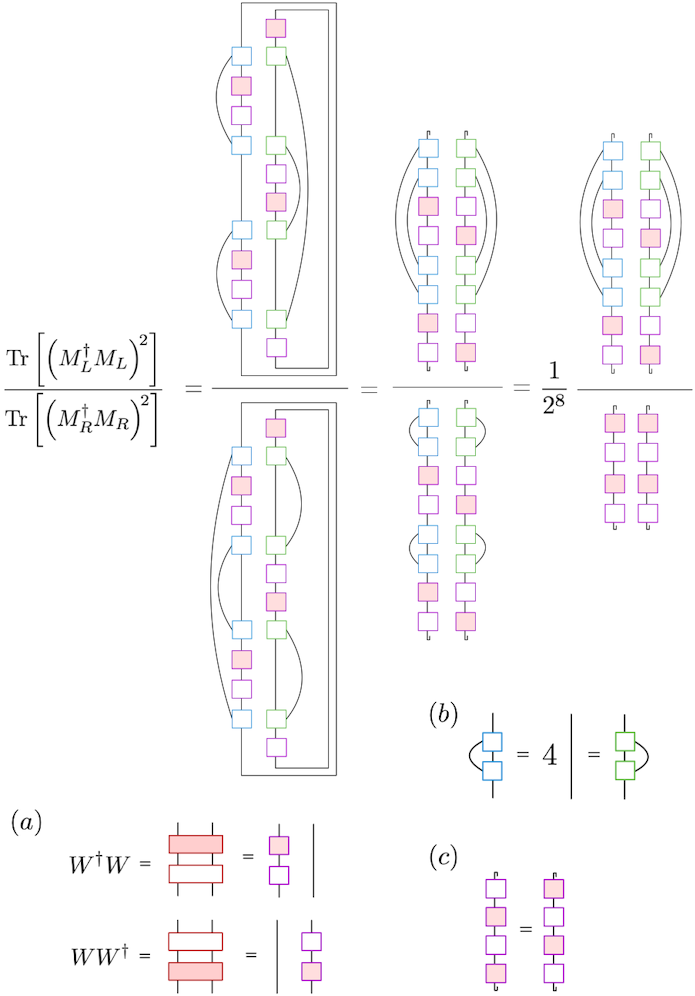}
    \caption{Argument of the current for the reset-swap QCA. The derivation follows the identities illustrated in the items below:
    (a) The products of two-site tensors,
    $W^\dagger W = R_1^\dagger\cdot \SWAP_{1,2} \cdot \SWAP_{1,2}\cdot R_1=R_1^\dagger R_1$ 
    and 
    $W^\dagger W = R_1^\dagger \cdot\SWAP_{1,2}\cdot \SWAP_{1,2}\cdot R_1=R_2 R_2^\dagger$,
    are separable because of the two arising consecutive swap operations.
    (b) The summed product of the tensors $A,B$ are trivial as these are here given by a linear combination of Pauli operators $\frac{1}{4}\sum_{a,b=0}^3 (\sigma^{a}\otimes\sigma^{b})(\sigma^{a}\otimes\sigma^{b})
    = 4\,\mathds{1}$.
    (c) The cyclic property of the trace is used, $\Tr[R^\dagger R R^\dagger R]=\Tr[R R^\dagger R R^\dagger]$.
    }
    \label{fig:resetswapcurrent}
\efig

The invariance of $I$ under the blocking process is confirmed as $W^\dagger W$ is factorizable: $W^\dagger W=R_1^\dagger R_1$ due to the swap operation of the map ($R_1$ acts only on the first system, there is an implicit identity on the second system), see Fig.~\ref{fig:resetswapcurrent}(a), where
Fig.~\ref{fig:resetswapblocking} shows the diagrammatic proof. 

\bfig
    \includegraphics[width=.82\columnwidth]{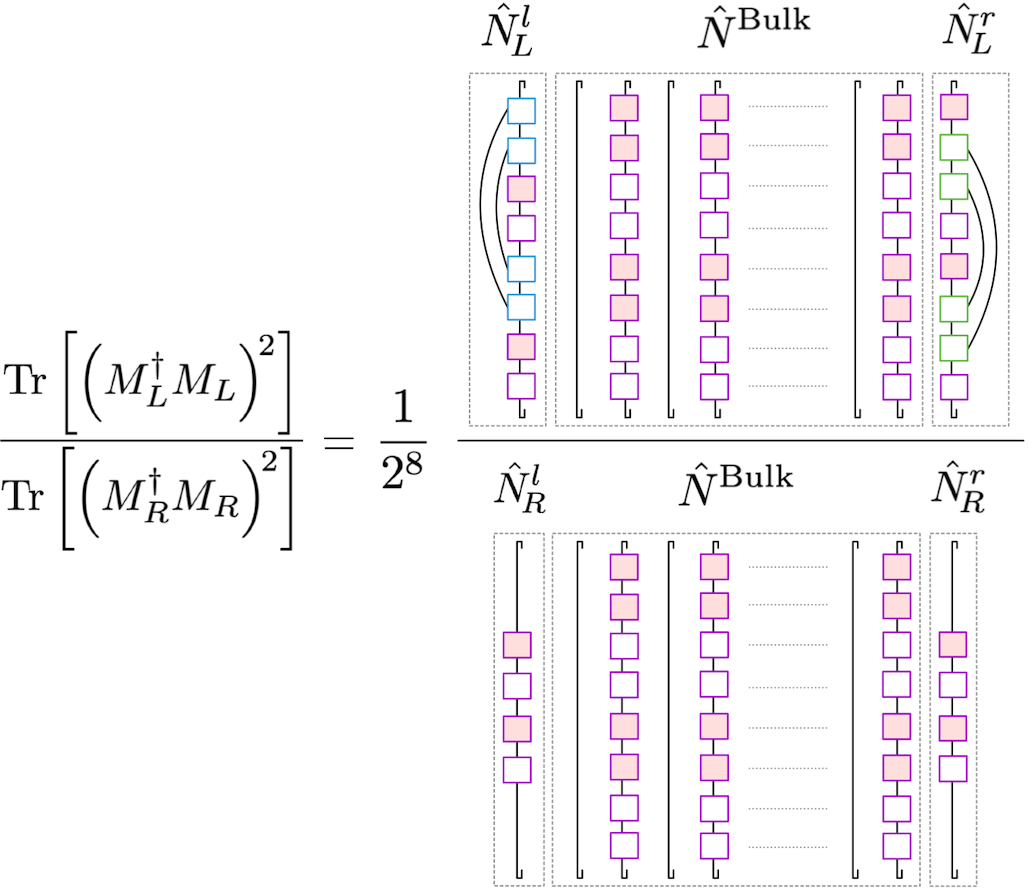}
    \caption{Invariance of the current under blocking for the coarse-grained reset-swap QCA, shown by the tensor network description of the argument of $I$.
    Note the marked terms $\hat N^\tx{Bulk}$ are the same for the numerator and the denominator and cancel out.
    The terms on the very left and right sites of the supercell, $\{\hat N_L^l,\hat N_L^r,\hat N_R^l,\hat N_R^r\}$ including the prefactor, equal the expression for the argument of $I$ associated with the non-coarse-grained QCA, see right-hand site in Fig.~\ref{fig:resetswapcurrent}.
    }
    \label{fig:resetswapblocking}
\efig

Next, Fig.~\ref{fig:resetswapcomposition} presents the non-additivity of the current under composition of two updates of the reset-swap QCA.
It is shown that the argument of $I$ without composition, 
$\frac{\Tr[\l(M_L^\dagger M_L\r)^{\!2}]}
{\Tr[\l(M_R^\dagger M_{_R}\r)^{\!2}]}$,
is a factor of the argument of $I$ with composition.
The current of the reset-swap map thus exhibits a recursive behavior under the composition of several updates of the QCA.

Another question to ask is how the composition of the reset-swap map with a unitary finite-depth circuit would change the current.
Intuitively, the information flow of the system should not change, as the current is vanishing for local unitary gates, as proved in \ref{sec:unitary}.
Therefore it is to highlight that the current does indeed not change under composition, or conjugation with a unitary, with a formal derivation presented in App.~\ref{AppResetswap}.
However, $I$ does change under composition with a unitary finite-depth circuit if two or more time steps of the QCA are composed. 
This is not surprising since the unitary operation can change the amount of information loss per time step.

\bfig
    \includegraphics[width=\columnwidth]{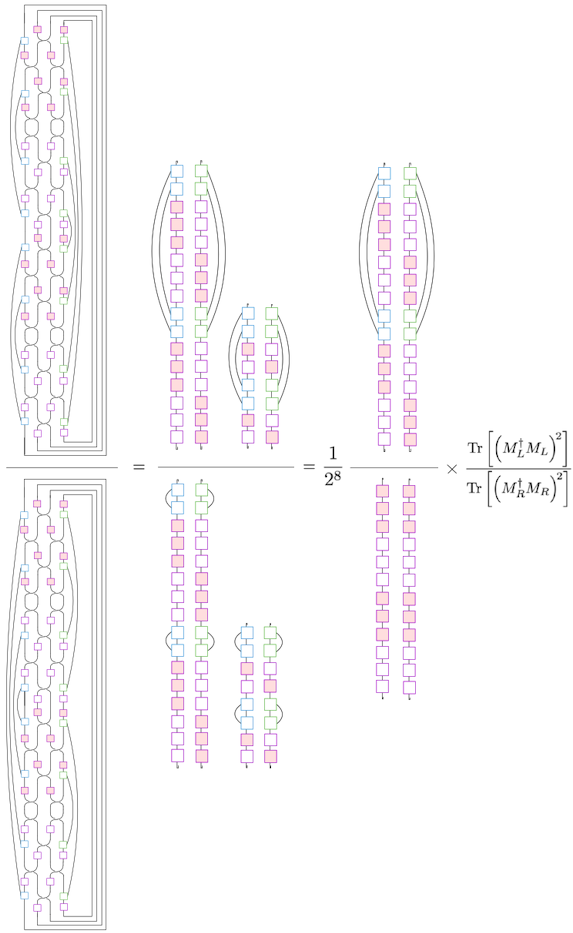}
    \caption{Argument of the current for two updates of the reset-swap QCA, including the blocking of two physical sites. The derivation is analogous to the derivation of the current without composition in Fig.~\ref{fig:resetswapcurrent}, but shows the swap operations in the first step (which are in Fig.~\ref{fig:resetswapcurrent} omitted for clarity).
    }
    \label{fig:resetswapcomposition}
\efig

Using the definitions in Eqs.~\eqref{eq:resetswapkraus} to \eqref{eq:resetswapM},
an analytical solution of the current $I$ as a function of $p$ can be derived:
\bs
\ba
    I 
    &= 2\, \log \l(
    \frac{\frac{1}{2}p^2-p+1}{p^2-p+1}
    \r),\label{eq:resetswap}
    \\
    I_\tx{c} &= 2I+\, 2\log \qty(  \xi  ) \tx{ with}\\
    \xi &= \frac{(p^2-p+1) \, ( p^4 - 4p^3 + 5p^2 - 2p + 1 ) }{p^6-6p^5+15p^4-19p^3+12p^2-3p+1
    } ,\label{eq:resetswapc}\nn
\ea
\label{eq:resetswapplot}
\es
\!where the second expression $I_\tx{c}$ corresponds to the current of the composition of two QCA updates. Note the current is only additive at the extremal points $p=0$ and $p=1$.

\bfig
    \centering
    \includegraphics[width=.9\columnwidth]{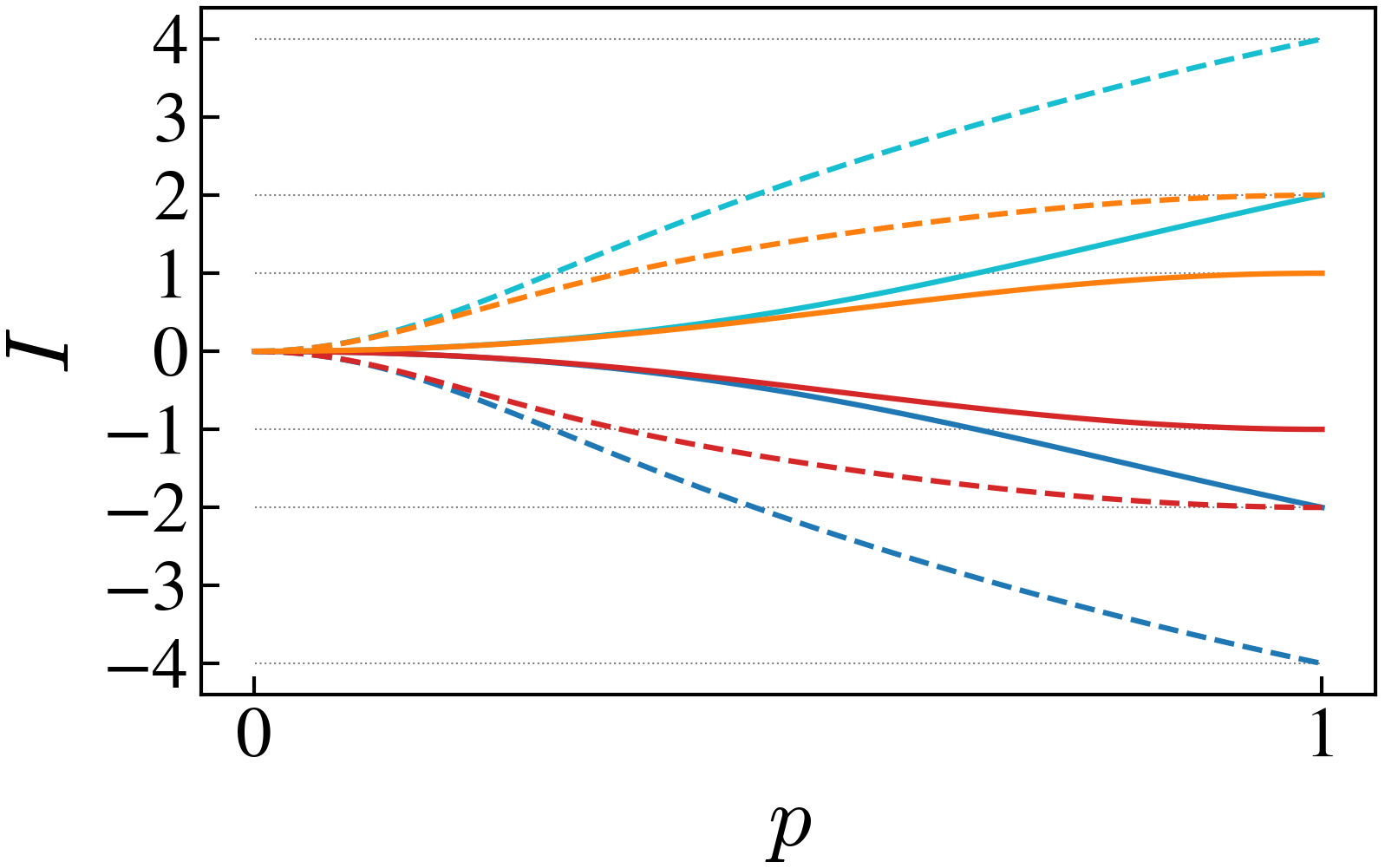}
    \caption{Information current $I$ for the reset-left-swap (blue), reset-right-swap (cyan), dephase-left-swap (red), and dephase-right-swap (orange) map as a function of the reset- or dephase rate $p$, where dashed lines indicate the composition of two time steps of the associated QCA.
    }
    \label{fig:resetswapplot}
\efig
The plot of the currents $I$ and $I_\tx{c}$ in Eq.~\eqref{eq:resetswapplot} are shown in Fig.~\ref{fig:resetswapplot} as functions of the reset-rate $p$.
One can see that the information currents for the reset-right swap and reset-left swap maps are equal and opposite. The current saturates at the same value as the index at $p=1$.
For the reset-right swap map, for $0\leq p<1$, $\Rank(M_L)=\Rank(M_R)=16$, however, at $p=1$, $\Rank(M_L)=1$ and $\Rank(M_R)=16$. This gives witness to the fact that the aforementioned index experiences a jump discontinuity from $\ind =0$ for $0\leq p<1$ to $\ind =2$ at $p=1$. Similar behavior occurs for the reset-left swap map but where at $p=1$, $\Rank(M_L)=16$ and $\Rank(M_R)=1$ and $\ind=-2$. A way to interpret this is that at $p=1$ four linearly independent operators are transported across a boundary in one direction, while only one, the $\dyad 0$ operator, is transported in the other direction.

\subsection{Dephase-swap map} \label{sec:dephaseswap}

The dephase-swap map is similar to the reset-swap map, where the local reset gate is replaced by a dephase gate, or equivalently exchanging local amplitude damping with phase damping. The Kraus operators describing the dephase operation are: 
\ba
K_0 &=\m(\sqrt{1-p/2}&0\\0&\sqrt{1-p/2}),\nn\\
K_1 &=\m(\sqrt{p/2}&0\\0&-\sqrt{p/2}),
\ea
with $p\in[0,1]$.
For this map, when acting on the left site for $0\leq p<1$, $\Rank(M_L)=\Rank(M_R)=16$, while for $p=1$, $\Rank(M_L)=4$ and $\Rank(M_R)=16$.
Hence, similar to the case for this reset-swap map, the index experiences a jump discontinuity from $\ind =0$ for $0\leq p<1$ to $\ind =1$ at $p=1$. Similar behavior occurs for the reset-left swap map but where at $p=1$, $\Rank(M_L)=16$ and $\Rank(M_R)=4$ and $\ind =-1$. Like the reset-swap map, at $p=1$, in one direction four linearly independent operators are transported across a boundary, but in contrast, two operators, $\dyad{0}$ and $\dyad{1}$, are transported in the other direction.

The solution of $I$ for the dephase-left swap map without and with composition (and blocking) of two QCA updates:
\bs
\ba
    I 
    &= \, \log \l(
    \frac{2 \left(\frac{1}{2} p^{2} - p + 1\right)^{2}}{\left(p^4 - 4 p^3 + 6 p^2 - 4 p + 2\right)}
    \r),
    \\
    I_\tx{c} &= 2I+\, 2\log \qty(  \xi  ) \tx{ with}\\
    \xi &= \frac{ (p^4 - 4 p^3 + 5 p^2 - 2 p + 1)^2}
    { p^8 - 8 p^7 + 28 p^6 - 56 p^5 + 69 p^4 - 52 p^3 + 22 p^2 - 4 p + 1} ,\nn
\ea
\es
where the corresponding plots are presented in Fig.~\ref{fig:resetswapplot}. Again, the current is only additive at the points $p=0,1$.
The information currents for the dephase-right swap and dephase-left swap maps are equal and opposite, whereby the current saturates at the same value as the index at $p=1$.\\

\subsection{Directed amplitude damping map}\label{sec:amplitudedamping}
In this section, the information flow of a QCA is investigated whose local tensors do not separate like $W^\dagger W = \tilde W_1 \otimes \tilde{\tilde W}_2$, where $\tilde W_1, \tilde{\tilde W}_2$ are arbitrary tensors acting on the associated left or the right site of a two-cell subsystem.

The quantum channel is described by a directed amplitude damping map from the $\ket{01}$ to the $\ket{00}$ state with probability $p\in[0,1]$:
\ba
    \ket{00}_S\ket{0}_E&\rightarrow \ket{00}_S\ket{0}_E\nn\\
    \ket{01}_S\ket{0}_E&\rightarrow \sqrt{1-p}\ket{01}_S\ket{0}_E + \sqrt{p}\ket{00}_S\ket{1}_E\nn\\
    \ket{10}_S\ket{0}_E&\rightarrow \ket{10}_S\ket{0}_E\nn\\
    \ket{11}_S\ket{0}_E&\rightarrow \ket{11}_S\ket{0}_E,\nn
\ea
where the subscripts $S$ and $E$ indicate the system or the environmental subsystem, respectively.
Tracing out the environment $E$ results in a quantum channel acting on (solely) the system $S$ defined by the Kraus operators
\ba
    K_0=\m(1 & 0 & 0 & 0 \\0 & \sqrt{1-p} & 0 & 0 \\0 & 0 & 1 & 0 \\0 & 0 & 0 & 1), \
    K_1=\m(0 & \sqrt{p} & 0 & 0 \\0 & 0 & 0 & 0 \\0 & 0 & 0 & 0 \\0 & 0 & 0 & 0),
\ea
which describe discrete-time dynamics.
These Kraus operators define the CPTP map in the considered vectorization formalism via 
\footnote{The definition of the map in its vectorized form follows from Eq.~(70) in \cite{gilchrist2011vectorization}. The conjugation of the tensors on the right-hand side of the tensor product can be omitted because $\{K_\mu\}_{\mu=1,2}$ are in this case real.
\ba
    V = \sum_{\mu=0,1}K_\mu\otimes K^*_\mu.
    \label{eq:K}
\ea
}

The channel could alternatively be generated by the continuous-time Lindblad dynamics described by the Master equation
\ba
    \rho(t+\tau)=e^{\mathcal{L}\tau}[\rho(t)],
\ea
which is defined by the Liouvillian
\ba
    \mathcal{L}[\rho] = \qty(L\rho L^\dagger - \frac{1}{2}\qty(L^\dagger L \rho + \rho L^\dagger L))
\ea
with jump operator
\ba
    L = \sqrt{-\frac{1}{\tau} \ln(1-p)} \dyad{00}{01},
\ea
where the time of the action of the channel is set to $\tau=1$ for simulating one corresponding discrete time step.

For completeness, the map that describes directed amplitude damping in the opposite direction, from the $\ket{10}$ to the $\ket{00}$ state, is also considered:
\ba
    \ket{00}_S\ket{0}_E&\rightarrow \ket{00}_S\ket{0}_E\nn\\
    \ket{01}_S\ket{0}_E&\rightarrow \ket{01}_S\ket{0}_E\nn\\
    \ket{10}_S\ket{0}_E&\rightarrow \sqrt{1-p}\ket{10}_S\ket{0}_E + \sqrt{p}\ket{00}_S\ket{1}_E\nn\\
    \ket{11}_S\ket{0}_E&\rightarrow \ket{11}_S\ket{0}_E,\nn
\ea
where Kraus operators are given by
\ba
    K_0=\m(1 & 0 & 0 & 0 \\0 & 1 & 0 & 0 \\0 & 0 & \sqrt{1-p} & 0 \\0 & 0 & 0 & 1), \
    K_1=\m(0 & 0 & \sqrt{p} & 0 \\0 & 0 & 0 & 0 \\0 & 0 & 0 & 0 \\0 & 0 & 0 & 0),
\ea
with corresponding jump operator
\ba
    L = \sqrt{-\frac{1}{\tau} \ln(1-p)} \dyad{00}{10}.
\ea

\bfig
    \centering
    \includegraphics[width=.95\columnwidth]{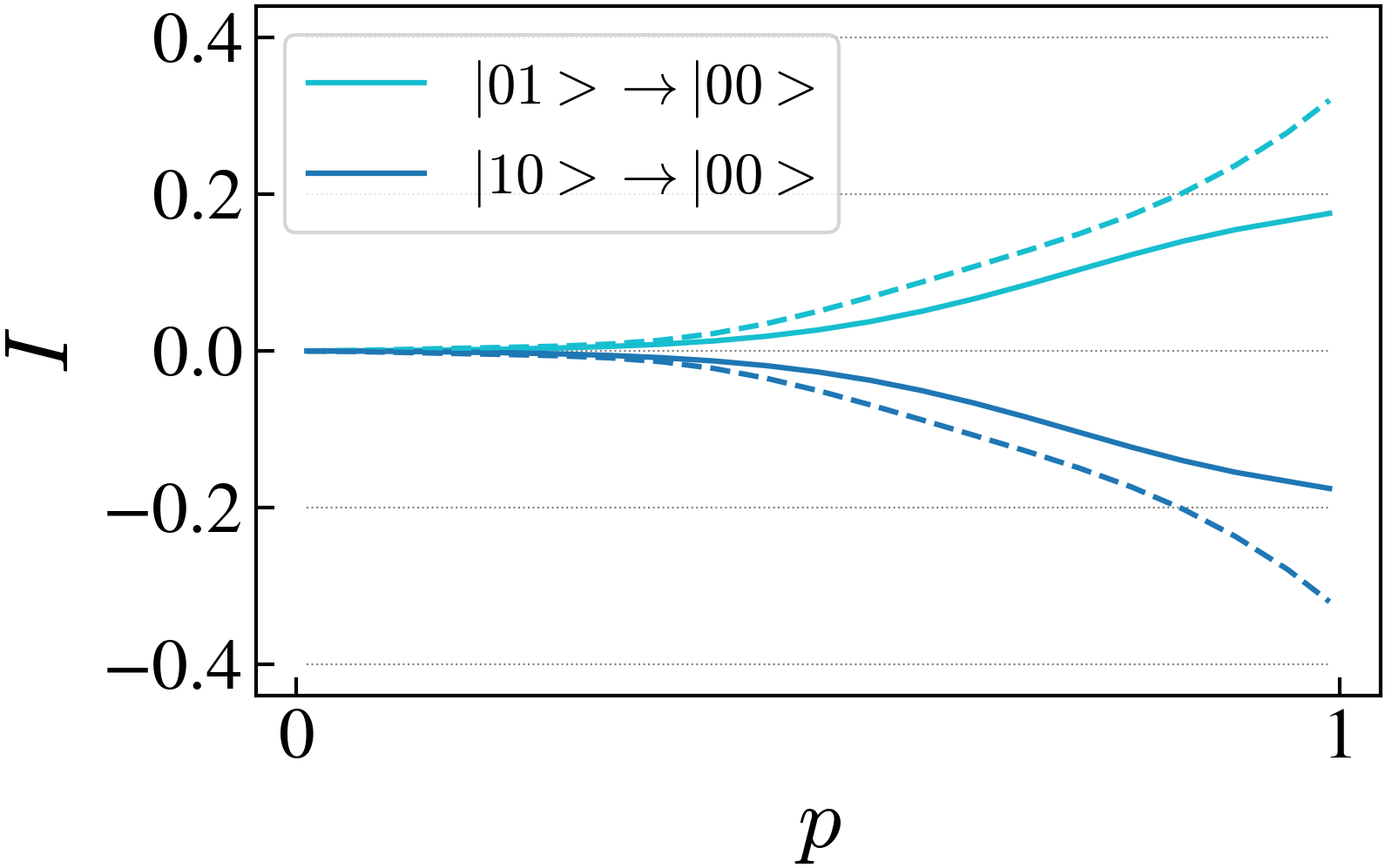}
    \caption{Information current $I$ for a QCA that describes directed amplitude damping as labeled, where $p$ is the amplitude damping rate. 
    Dashed lines correspond to the composition two time steps of the QCA.}
    \label{fig:amplitudedampingplot}
\efig
The resulting current is in Fig.~\ref{fig:amplitudedampingplot} plotted as a function of the damping rate $p\in[0,1]$:
As expected, the current increases monotonically with the damping rate, and
no additivity under composition is observed. Additionally, this is a first example we have described so far where $W^\dagger W$ is not separable, see proof in App.~\ref{sec:Appamplitudedamping}. The current is therefore not invariant under blocking.

\subsection{Asymmetric swap map} \label{sec:asymmetricswap}
An asymmetric swap operation acting on two physical sites is investigated next. 
The asymmetry is realized by considering a partial swap gate that is dependent on the initial state: The $\ket{01}$ state is ``swapped" onto the $\ket{10}$ state with a certain probability $p_{01}\in[0,1]$, where the latter is not necessarily equal to the transition probability $p_{10}\in[0,1]$ from the $\ket{10}$ to the $\ket{01}$ state; see below definition of the channel.
\ba
    \ket{00}_S\ket{0}_E&\rightarrow \ket{00}_S\ket{0}_E\nn\\
    \ket{01}_S\ket{0}_E&\rightarrow \sqrt{1-p_{01}}\ket{01}_S\ket{0}_E + \sqrt{p_{01}}\ket{10}_S\ket{1}_E\nn\\
    \ket{10}_S\ket{0}_E&\rightarrow \sqrt{1-p_{10}}\ket{10}_S\ket{0}_E + \sqrt{p_{10}}\ket{01}_S\ket{1}_E\nn\\
    \ket{11}_S\ket{0}_E&\rightarrow \ket{11}_S\ket{0}_E\nn
\ea
By tracing out the environmental subsystem, a quantum channel acting on the system can be defined by the Kraus operators
\ba
    K_0&=\m(1 & 0 & 0 & 0 \\0 & \sqrt{1-p_{01}} & 0 & 0 \\0 & 0 & \sqrt{1-p_{10}} & 0 \\0 & 0 & 0 & 1), \nn\\
    K_1&=\m(0 & 0 & 0 & 0 \\0 & 0 & \sqrt{p_{10}} & 0 \\0 & \sqrt{p_{01}} & 0 & 0 \\0 & 0 & 0 & 0),
\ea
that define the (vectorized) superoperator using Eq.~\eqref{eq:K}.
The continuous-time version of this map is generated by the Liouvillian
\ba
    \mathcal{L}[\rho] = \sum_{k=1,2}\qty(L^{(k)}\rho L^{(k)^\dagger} - \frac{1}{2}\qty(L^{(k)^\dagger} L^{(k)} \rho + \rho L^{(k)^\dagger} L^{(k)})),
\ea
which includes two jump operators
\ba
    L^{(1)} &= \sqrt{-\frac{1}{\tau} \ln(1-p_{01})} \dyad{10}{01}, \nn\\
    L^{(2)} &= \sqrt{-\frac{1}{\tau} \ln(1-p_{10})} \dyad{01}{10},
\ea
that are each associated with the damping of the $\ket{01}$ or the $\ket{10}$ state, respectively.

In total, no information flow is observed, $I=0$, despite the asymmetry of the swap operation for $p_{01}\neq p_{10}$.
This meets expectations nonetheless because 
of the swap-symmetry of the map;
under conjugation with the unitary $X_1 X_2$,
the jump operators satisfy $\forall \ p_{01}\neq p_{10}$:
\ba
    \SWAP_{1,2}\,(X_1 X_2)\, L^{(1)}\, (X_1 X_2)\, \SWAP_{1,2} &= L^{(1)}, \nn\\
    \SWAP_{1,2}\, (X_1 X_2)\, L^{(2)}\, (X_1 X_2) \, \SWAP_{1,2} &= L^{(2)}.
\ea
Note that the conjugation with a unitary can be applied w.l.o.g.~as it does not change the current.
The proof  of the invariance under the swap operation is depicted in Fig.~\ref{fig:swapsymmetricproof}.

\subsection{Integrable non-unitary maps} \label{sec:integrable}
 Recently there has been an investigation of one-dimensional integrable non-unitary QCA \cite{integrable21,integrable22}. These are models where the open systems dynamics are of Lindblad type and the maps generated by each Lindblad summand define a non-unitary map that is a solution to the Yang-Baxter equation. 

A classification of integrable Lindbladians are given in \cite{integrable21} and their corresponding CPTP maps are presented in \cite{integrable22}. All their listed models (A1, A2, B1, B2, B3) are specified below in terms of their CPTP map representation, and the corresponding information currents $I$ are inspected.

The models studied in \ref{tab:integrable} come in two classes. The models of {\rm A} type have Lindblad jump operators which are lower triangular with at most two elements below the diagonal in the logical basis, and they are non-unital maps. Model A1 has one element below the diagonal, and model A2 has two elements below the diagonal.  The models of ${\rm B}$ type conserve total spin projection $S^z=\frac{1}{2}\sum_j Z_j$ and only models B1 and B2 are unital.
A summary of the properties of these models is give in Tab.~\ref{tab:integrable}.
\begin{table}[h]
\begingroup
\setlength{\tabcolsep}{10pt} 
\renewcommand{\arraystretch}{1.5} 
\begin{center}
\begin{tabular}{ c| c c c c c }
       & A1 & A2 & B1 & B2 & B3 \\ \hline
unital & \xmark & \xmark & \cmark & \cmark & \xmark \\
$\Delta S^z=0$ & \xmark & \xmark & \cmark & \cmark & \cmark \\
lower triangular & \cmark & \cmark & \xmark & \xmark & \xmark \\
$|I|>0$   & \cmark & \cmark & \xmark & \cmark & \xmark
\end{tabular}
\end{center}
\caption{Overview of the properties of the integrable models A1, A2, B1, B2, and B3 presented in \cite{integrable22}. The CPTP maps have been formulated and categorized in \cite{integrable22} according to whether they are unital, lower triangular and/or conserve $S^z$. Three out of the five models are in this work shown to have a non-zero information current $|I|$ – notably independent of the unital property, the conservation of $S^z$ and whether the superopartor is lower triangular.}
\label{tab:integrable}
\endgroup
\end{table}

\subsubsection{Models A1 and A2}\label{sec:A}
The CPTP maps of models A1 and A2 in \cite{integrable22} are investigated at first. 
Model A1 is in the Kraus representation given by the two Kraus operators
\bs
\ba
    K_0&=e^{-u/2}\m(
    1 & 0 & 0 & 0 \\
    0 & e^{-u/2} & 0 & 0 \\
    0 & 0 & e^{u/2} & 0 \\
    0 & 0 & 0 & 1), \\
    K_1&=\sqrt{1-e^{-u}}\,\m(
    1 & 0 & 0 & 0 \\
    0 & e^{-u/2} & 0 & 0 \\
    0 & -ie^{i\phi} & 0 & 0 \\
    0 & 0 & 0 & 1),
\ea
\label{eq:krausA1}
\es
\!\!where $0\leq u\in\mathbb{R}$  and $\phi\in\mathbb{R}$ represents an arbitrary phase factor. (Due to the imaginary phase $-ie^{i\phi}$ in $K_1$ this is in this work the first presented map that includes complex numbers.)

Model A2 is defined by the Kraus operators
\bs
\ba
    K_0&=e^{-u}\m(
    1 & 0 & 0 & 0 \\
    0 & e^{-u/2} & 0 & 0 \\
    0 & -\tau(e^u-1) & e^{u/2} & 0 \\
    0 & 0 & 0 & e^u), \\
    K_1&=\sqrt{e^u-1}\;e^{-u}\m(
    1 & 0 & 0 & 0 \\
    0 & 0 & 0 & 0 \\
    0 & \tau & e^{u/2} & 0 \\
    e^{u/2} & 0 & 0 & 0),
\ea
\label{eq:krausA2}
\es
where $\tau=\pm1$. The corresponding vectorized CPTP map representations are (as in the previous examples) given by Eq.~\eqref{eq:K}.

\bfig
    \centering
    \includegraphics[width=.95\columnwidth]{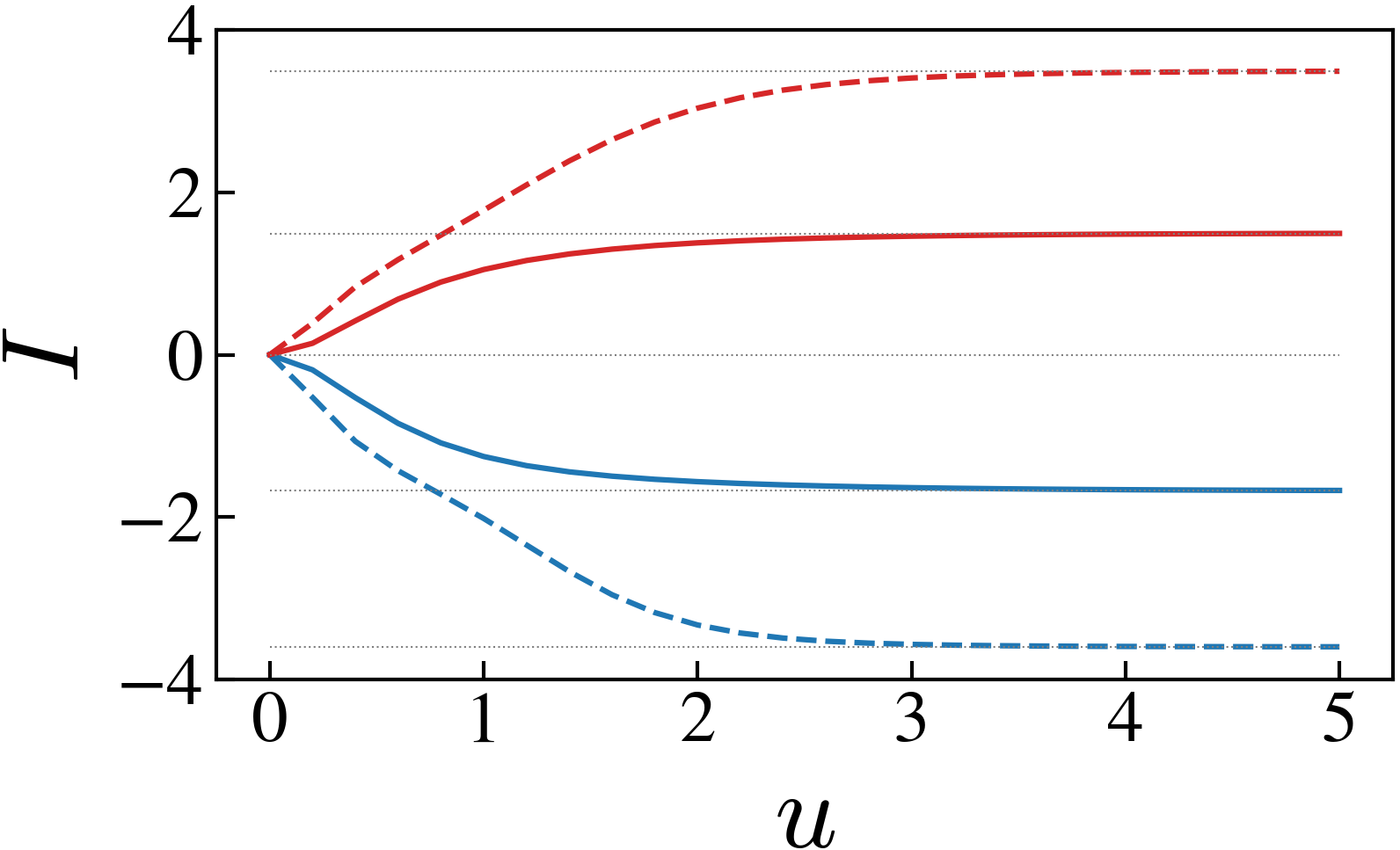}
    \caption{Information current $I$ corresponding to the integrable channels given in models A1 (blue) and A2 (red) in \cite{integrable22}, see Kraus operators in Eqs.~\eqref{eq:krausA1} and \eqref{eq:krausA2}, respectively. Note that in calculations of map A1, the choice of the phase factor $\phi\in\mathbb{R}$ does not change $I$, while for model A2, the current does analogously turn out to be independent of the choice of the sign of $\tau=\pm1$.
    The current $I_\tx{c}$ associated with the composition of two time steps of the corresponding QCA is represented by dashed lines in the same color as the model.}
    \label{fig:ybeA}
\efig

The associated currents $I$ are plotted as a function of $u\geq0$ in Fig.~\ref{fig:ybeA}.
One can see that for both maps $|I|$ is monotonically increasing with $u$ and saturates at about $u=3$ to a constant value: $I=-1.68$ and $I_\tx{c}=-3.61$ for model A1, and $I=1.50$ and $I_\tx{c}=3.49$ for model A2.


\subsubsection{Model B1}\label{sec:B1}
The CPTP map of model B1 in \cite{integrable22} is expressed by
\ba
    K_0=
    \sqrt{\frac{1}{u+1}}\;
    \mathds{1}_{4\times4},\
    K_1=\sqrt{\frac{u}{u+1}}\,\m(
    1 & 0 & 0 & 0 \\
    0 & 0 & \tau & 0 \\
    0 & \tau & 0 & 0 \\
    0 & 0 & 0 & \kappa\tau),
    \label{eq:krausB1}
\ea
where $u\geq0$, $\tau,\kappa=\pm1$, and $\mathds{1}_{4\times4}$ represents the identity channel acting on two qubits.

Because of the swap symmetry of this channel \eqref{eq:krausB1}, the current is vanishing, $I=0$; see proof in Sec.~\ref{sec:swapsymmetric}.

\subsubsection{Model B2}\label{sec:B2}
Model B2 is defined by the Kraus operators
\ba
    K_0 = \sqrt{\beta}\, G(u-v),\
    K_1 = \sqrt{\alpha}\, G(u+v)\, (Z\otimes\mathds{1}_{2\times2}),
    \label{eq:krausB2}
\ea
where
\ba
    &G(u)=
    \frac{1}{\cosh(u)}\nn\\
    &\times\m(
    \cosh(u) & 0 & 0 & 0\\
     0 & 1 & -i\sinh(u) e^{i\phi} & 0 \\
     0 & -i\sinh(u) e^{-i\phi} & 1 & 0 \\
     0 & 0 & 0 & \cosh(u)),
\ea
$\phi\in[0,2\pi)$ is an arbitrary phase, and $\mathds{1}_{2\times2}$ labels the identity channel acting on one qubit. The normalization constants $\sqrt{\beta}$ and $\sqrt{\alpha}$ of each, $K_0$ or $K_1$, respectively, are given by
\ba
    \beta = 
    \frac{\cosh(u-v)\cosh(\xi-\eta)}{\cosh(u-v)\cosh(\xi-\eta)+\cosh(u+v)\sinh(\xi-\eta)}
    \label{eq:beta}
\ea
and $\alpha = 1-\beta$ under the conditions that $u\geq v$ and
\ba
    \frac{\sinh(2\xi)}{\sinh(2u)}
    =\frac{\sinh(2\eta)}{\sinh(2v)}.
    \label{eq:xieta}
\ea

The associated current is shown below in Fig.~\ref{fig:ybeB2}.

\bfig
    \centering
    \includegraphics[width=.95\columnwidth]{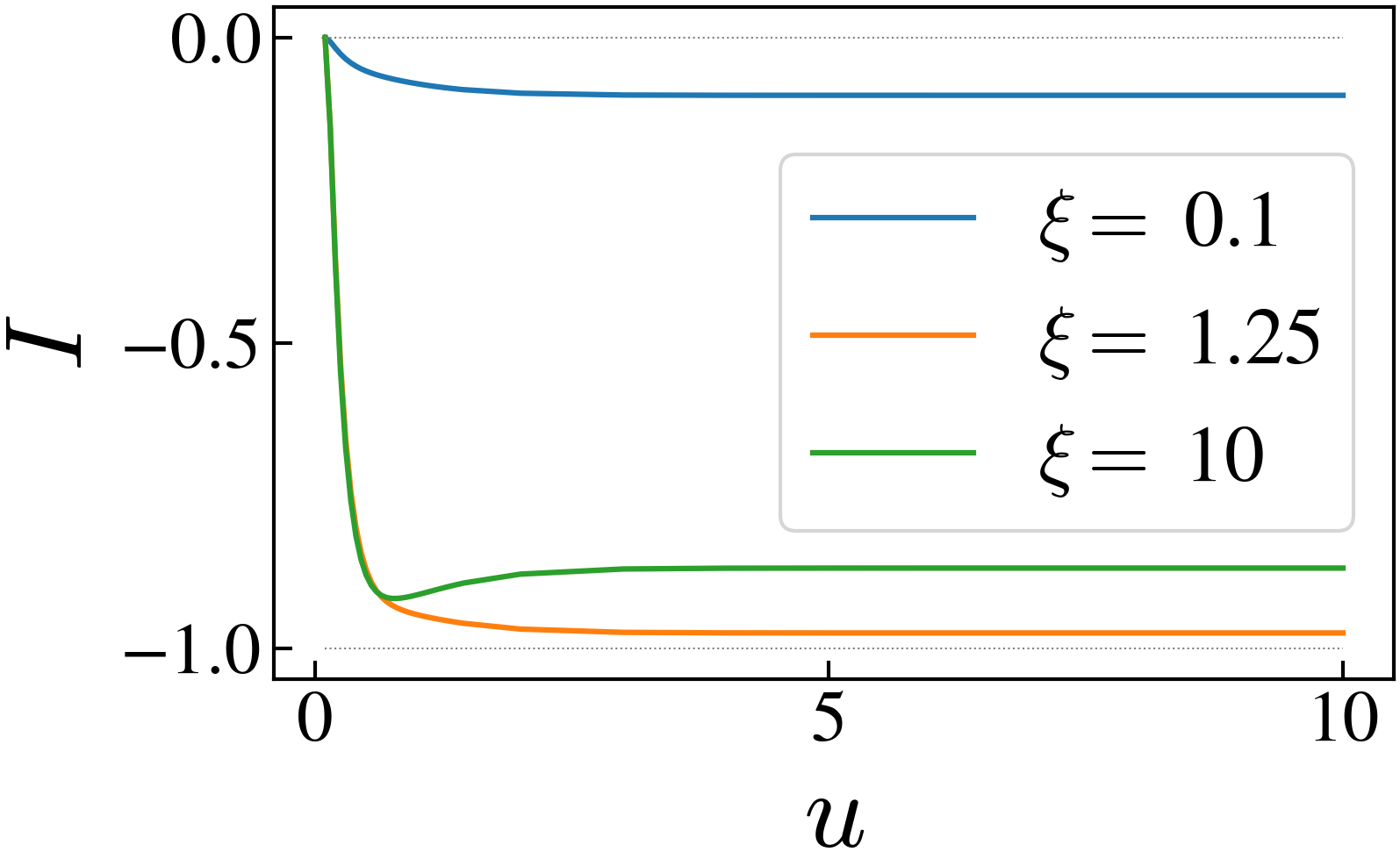}
    \caption{Information current $I$ for the integrable CPTP map given in model B2 in \cite{integrable22}. The map is given by the Kraus operators in Eq.~\eqref{eq:krausB2}, where for the presented calculations the parameters $v=0.1$ and $\phi=0$ are chosen. Due to the above condition $u\geq v$, the plot starts at $u=v=0.1$.
    }
    \label{fig:ybeB2}
\efig
Note that $I=0$ if $u=v=0.1$ due to the unitarity of the map is in this case.
For all $u\geq v$, $I$ is decreasing (monotonically for small $\xi$) and saturates to a constant value for all about $u\geq3$: $I=-0.095$ for $\xi=0.1$, $I=-0.97$ for $\xi=1.25$, and $I=-0.87$ for $\xi=10$. The constant $\xi=1.25$ has been chosen because the absolute value of the current is for this value maximum, given $v=0.1$.

For $u\rightarrow\infty$ the current is observed to be vanishing due to the unitarity of the map in this limit.
If one were to set $\xi=0$, then one would find with Eq.~\eqref{eq:xieta} that $\xi=\eta$, leading to $\beta=1$ with Eq.~\eqref{eq:beta} and $\alpha=1-\beta=0$, such that $K_0=G(u-v)$ and $K_1=0$ according to Eq.~\eqref{eq:krausB2} --- the channel is therefore unitary and $I$ vanishes for $\xi=0$.

It is notable that the computational complexity of calculating the current for this model is drastically increased in comparison to the previously presented models. Here, the ranks of the considered local tensors are equal to 16, or $16^2=256$, for one or two composed time steps, respectively --- in comparison to each 4 or 16 in both models A1 and A2 in Sec.~\ref{sec:A}. Multiplying the rank, which is effectively the size of the virtual degree of freedom of the MPO tensors, with the dimension of the (vectorized) physical degree of freedom, $4^2=16$ for two lattice sites or $4^4=256$ for four lattice sites, one obtains the size of the matrices $M_L$ and $M_R$: $(256\times256)$ for a single time step, or $(65\,536\times65\,536)$, i.e.~squared, for two composed updates of the associated QCA. Because the calculation of the information current $I$ involves more complex matrix transformations that are computationally more costly for matrices of the latter size, additional plots for the current $I_\tx{c}$ of two QCA time steps are not performed here. However, the value $I_\tx{c}(u=5,v=0.1,\xi=1.25,\phi=0)=-1.949669$ has been calculated, which, interestingly, is very close (up to an error of $10^{-5}$) equal to twice the current without the composition: $2I(u=5,v=0.1,\xi=1.25,\phi=0)=-1.949698$ \footnote{The scaling factor could be a result of $I$ not being fully saturated for $u=5$, see plot in Fig.~\ref{fig:ybeB2}, but where we leave this question for further research.}. This shows that the net information flow would linearly increase as a function of time.

\subsubsection{Model B3}\label{sec:B3}

Model B3 is given by the Kraus operators
\bs
\ba
    &K_0=\nn\\
    &\frac{1}{\eta}\m(
    \eta & 0 & 0 & 0 \\
    0 & 1 & -i(1-\gamma)\sinh(\alpha)e^{i\phi}  & 0 \\
    0 & -i(1+\gamma)\sinh(\alpha)e^{-i\phi} & 1 & 0 \\
    0 & 0 & 0 & \eta), \\
    &K_1=\frac{\xi}{\eta}\m(
    0 & 0 & 0 & 0 \\
    0 & -e^{\alpha}(1-\gamma) & i(1-\gamma)e^{i\phi} & 0 \\
    0 & i(1+\gamma)e^{-i\phi} & e^{\alpha}(1+\gamma) & 0 \\
    0 & 0 & 0 & 0),
\ea
\es
where $\gamma\geq0$, $\alpha=(\gamma^2+1)u/2$, $\eta=\gamma\sinh(\alpha)+\cosh(\alpha)$, and $\xi=\sqrt{\gamma\sinh(\alpha)/[(1+\gamma^2)\cosh(\alpha)+2\gamma\sinh(\alpha)]}$.

The corresponding current has been found to be vanishing for all $u\geq0$ and $\gamma\geq0$. 
\\

All in all, the presented integrable non-unitary models might prove to be of fundamental interest in relation to the information current for further analysis. Because all CPTP maps can be classified using the proposed measure, notably independent of the type of particle or the choice of operator basis, our information flow measure e.g.~generalizes the particle current $J_k$ for model B2 in \cite{integrable21},  Eq.~(20), which has only been defined for states in the particle number basis.

%
%

\section{Conclusion}
We have introduced a measure of information flow for open quantum system dynamics which captures features that are not obtained from the unitary case.
Considering the MPO representation of QCA, the information current has been constructed using the left or right partitionings of the local tensors of the MPO, similar to the description of the index theory using MPUs \cite{PhysRevB.98.245122,Ignacio_Cirac_2017}. Here, we have rewritten the index in terms of the difference of R\'enyi-0 entropies of the singular values associated with the left and right partitionings of these tensors. The current, in contrast, has been for non-unitary systems defined by the difference in \Renyitwo entropies of the corresponding singular values.

The current is locally computable, vanishing for finite-depth unitary circuits and SWAP-symmetric QCA, and is continuous in the noise parameters for maps.
It has been shown to exhibit the same value as the index for the shift map, the standard example of a non-local unitary QCA.

In contrast to the index, the information current is not generically invariant under blocking of lattice sites, nor is it additive under composition of QCA updates.
Failure to be additive under composition is not surprising given the nonlinearity of the function we are evaluating. 
The general lack of invariance under blocking indicates that bulk properties of the (blocked) supercells can change the information flow at the edges of the cell, though there is a large class of open systems dynamics where the information current is invariant under blocking.

A particularly interesting use case for our measure are the integrable models in Sec.~\ref{sec:integrable}, which have been shown to exhibit a particle flow in certain cases; see e.q.~Eq.~(20) in \cite{integrable21}. As the information current $I$ is independent of the type of particle and the choice of basis, our measure provides a generalization to the notion of a particle flow in \cite{integrable21}.

One might ask whether the information current for non-unitary QCA has a thermodynamic interpretation wherein a spatially periodic heat exchange between the environment and the system generates a net information current. For example, in the case of the reset-swap map, by Landauer's principle, the reset map necessarily involves heat exchange at every second site with the environment of $k_BT\ln 2$, where $k_B$ is the Boltzmann constant and $T$ is the temperature of the environment. We leave this matter to future work.

Lastly, it is noted that the implementation of non-unitary QCA can be realized using lattices of ultracold atoms excited to Rydberg states \cite{2020_whitlock}. Radius one rules in this implementation can be used for a variety of tasks including dissipative entangled resource state preparation and the study of non-equilibrium phase transitions \cite{2021_nigmatullin,Lesanovsky_DarkPhaseTransitions,Lesanovsky_NonequilibriumPhaseTransitions}.
The information current could thereby be computed indirectly  using process tomography on varying inputs states in the neighborhood of the QCA rule to determine the positive Hermitian operators 
$\sigma_{\beta}=M^{\dagger}_{\beta}M_{\beta}/\Tr[M^{\dagger}_{\beta}M_{\beta}]\ \forall\,\beta\in\{L,R\}$ in Eq.~\eqref{eq:sigmaop}.

\section{Acknowledgments}
We acknowledge helpful discussions with David Gross. This research was supported by the Australian Research Council Centre of Excellence for Engineered Quantum Systems (EQUS, CE170100009),
by the Sydney Quantum Academy, Sydney, NSW, Australia,
and with the assistance of resources from the National Computational Infrastructure (NCI), which is supported by the Australian Government.

\bibliography{biblio}

\appendix

\section{Singular values of \texorpdfstring{$M_L$}{TEXT} and \texorpdfstring{$M_R$}{TEXT} of local unitary QCA}
\label{AppA}
Here we prove that the singular values of $M_L$ an $M_R$ are equal for local unitary QCA. The contractions giving the traces of the second moments of $M_L^{\dagger}M_L$ and $M_R^{\dagger}M_R$ are shown in Fig.~\ref{fig:unitaryproof}. If the QCA rule is unitary, then $W^{\dagger}W=\mathds{1}$, and as is apparent from the diagram,
\bs
\ba
    &\Tr[ \l(M_L^{\dagger}M_L\r)^{\!2}] \nn\\
    &\qquad =
    \Tr[\l(\sum_{k=1}^{\chi} A^{\dagger}_{k}A_{k}\r)^{\!\!2}]\Tr[\l(\sum_{k=1}^{\chi} B_{k}B^{\dagger}_{k}\r)^{\!\!2}],
\ea
\ba
    &\Tr[\l(M_R^{\dagger}M_{_R}\r)^{\!2}] \nn\\
    &\qquad =
    \Tr[\l(\sum_{k=1}^{\chi} A_{k}A^{\dagger}_k\r)^{\!\!2}]\Tr[\l(\sum_{k=1}^{\chi} B^{\dagger}_{k}B_{k}\r)^{\!\!2}].
\ea
\label{Tracemoments}
\es

For unitary QCA, the tensor $V$ is also unitary.
From Eq.~\eqref{localtensors}, the local tensors are
\ba
    B_k=\sum_{r=1}^{d^2} \sqrt{D_{k,k}} Y_{r,k} \hat O_r, \
    A_k=\sum_{s=1}^{d^2} \sqrt{D_{k,k}} X^{\dagger}_{k,s}\hat O_s.
\ea
Using the fact that $V^{\dagger}V=\mathds{1}\otimes \mathds{1}$ we have
\ba
    \mathds{1}\otimes \mathds{1}
    &= 
    \sum_{k,k'=1}^{\chi} B^{\dagger}_kB_{k'}\otimes A^{\dagger}_kA_{k'} 
    \nn\\
    &=\sum_{k,k'=1}^{\chi} D_{k,k}D_{k',k'}
    \nn\\&\qquad
    \times\Bigg(\sum_{r,r'=1}^{d^2}Y^{\ast}_{r,k}Y_{r',k'}O^{\dagger}_r O_{r'}
    \nn\\&\qquad\qquad
    \otimes
    \sum_{s,s'=1}^{d^2}(X^{\dagger})^{\ast}_{k,s}X^{\dagger}_{k',s'}O^{\dagger}_s O_{s'}\Bigg).
\label{firsteq}
\ea

Taking a partial trace over the second factor, i.e.~on Hilbert space $\mathcal{H}_2\times \mathcal{H}_2^{\ast}$, gives
\ba
    d^2 \mathds{1}
    &= \sum_{k,k'=1}^{\chi} D_{k,k}D_{k',k'} \sum_{r,r'=1}^{d^2}Y^{\ast}_{r,k}Y_{r',k'}O^{\dagger}_r O_{r'} \nn\\&\qquad\qquad
    \sum_{s,s'=1}^{d^2}(X^{\dagger})^{\ast}_{k,s}X^{\dagger}_{k',s'}\Tr[O^{\dagger}_s O_{s'}] \nn\\
    &= \sum_{k,k'=1}^{\chi} D_{k,k}D_{k',k'} \sum_{r,r'=1}^{d^2}Y^{\ast}_{r,k}Y_{r',k'}O^{\dagger}_r O_{r'} \nn\\&\qquad\qquad
    \sum_{s,s'=1}^{d^2}X^{\dagger}_{k',s'}X_{s,k}\delta_{s,s'} \nn\\
    &= \sum_{k=1}^{\chi} D_{k,k}^2 \sum_{r,r'=1}^{d^2}Y^{\ast}_{r,k}Y_{r',k}O^{\dagger}_r O_{r'} \nn\\
    &= \sum_{k=1}^{\chi} D_{k,k} B^{\dagger}_k B_k.
\label{secondeq}
\ea
Taking a second trace gives
\ba
    \sum_{k=1}^{\chi} D_{k,k}^2=d^4.
\ea
Similarly taking the partial trace on the first factor, i.e.~on Hilbert space $\mathcal{H}_1\times \mathcal{H}_1^{\ast}$, gives
\ba
    d^2 \mathds{1}=\sum_{k=1}^{\chi} D_{k,k} A^{\dagger}_k A_k.
\ea

An analogous argument using  $VV^{\dagger}=\mathds{1}$ shows that 
\ba
    \sum_{k=1}^{\chi} D_{k,k} B_k B^{\dagger}_k= \sum_{k=1}^{\chi} D_{k,k} A_k  A^{\dagger}_k=d^2\mathds{1}.
\label{thirdeq}
\ea
Additionally, 
\ba
    \Tr[\sum_{k=1}^{\chi}A^{\dagger}_kA_k]
    &=\sum_{k=1}^{\chi}\sum_{s,s'=1}^{d^2}D_{k,k}(X^{\dagger})^{\ast}_{k,s}X^{\dagger}_{k,s'}\Tr[\hat O^{\dagger}_s \hat O_{s'}] \nn\\
    &=\sum_{k=1}^{\chi}D_{k,k}\sum_{s=1}^{d^2}X^{\dagger}_{k,s}X_{s,k}\nn\\
    &=\Tr[D],
    \label{traceeq}
\ea
and similarly, 
\ba
    \Tr[\sum_{k=1}^{\chi}B^{\dagger}_kB_k]
    =\Tr[\sum_{k=1}^{\chi}A_kA^{\dagger}_k]
    =\Tr[\sum_{k=1}^{\chi}B_kB^{\dagger}_k]
    =\Tr[D].
\ea

Now we can rewrite Eq.~\eqref{firsteq} based on $V^{\dagger}V=\mathds{1}\otimes \mathds{1}$ as
\ba
    &\sum_{k,k'=1}^{\chi} B^{\dagger}_kB_{k'}\otimes A^{\dagger}_kA_{k'}
    \nn\\&\qquad
    =\sum_{k=1}^{\chi} \frac{D_{k,k}}{d^2} B^{\dagger}_k B_k\otimes \sum_{k=1}^{\chi} \frac{D_{k,k}}{d^2} A^{\dagger}_k A_k
    =\mathds{1}\otimes \mathds{1}.
\ea
This can be viewed as a rank one operator singular value decomposition of the identity. We could have just as well defined local tensors $\tilde{B}_k=\sqrt{D_{k,k}}B_k,\tilde{A}_k=A_k/\sqrt{D_{k,k}}$, which also has a rank one singular value decomposition
\ba
    &\sum_{k,k'=1}^{\chi} \tilde{B}^{\dagger}_k\tilde{B}_{k'}\otimes \tilde{A}^{\dagger}_k\tilde{A}_{k'}\nn\\
    &\qquad=\sum_{k=1}^{\chi} \frac{D^2_{k,k}}{d^4} B^{\dagger}_k B_k\otimes \sum_{k=1}^{\chi} A^{\dagger}_k A_k
    =\frac{1}{c}\mathds{1}\otimes c\mathds{1},
\ea
where the constant $c=\Tr[D]/d^2$ is determined by Eq.~\eqref{traceeq}.
This implies $\sum_{k=1}^{\chi} A^{\dagger}_k A_k=c\mathds{1}$.
A similar result is found when weighting the local tensors the other way $\tilde{B}_k=B_k/\sqrt{D_{k,k}},\tilde{A}_k=A_k \sqrt{D_{k,k}}$ implying $\sum_{k=1}^{\chi} B^{\dagger}_k B_k=c\mathds{1}$.
Using the same argument but with $V V^{\dagger}=\mathds{1}\otimes \mathds{1}$ we find 
\ba
    \sum_{k=1}^{\chi} A_k A^{\dagger}_k=\sum_{k=1}^{\chi} B_k B^{\dagger}_k=\sum_{k=1}^{\chi} A^{\dagger}_k A_k=\sum_{k=1}^{\chi} B^{\dagger}_k B_k=c \mathds{1}.
    \label{eq:unitarycondition}
\ea

Returning to Eq.~\eqref{Tracemoments}, for local unitary QCA, the relevant terms are
\ba
    \Tr[(M_L^{\dagger}M_L)^m]
    &= \Tr[(\sum_{k} A^{\dagger}_{k}A_{k})^m]\Tr[(\sum_{k} B_{k}B^{\dagger}_{k})^m]\nn\\
    &= \Tr[(c \mathds{1})^m] \Tr[(c\mathds{1})^m]\nn\\
    &= d^4 (\Tr[D]/d^2)^{2m}\nn\\
    &= \Tr[(M_R^{\dagger}M_R)^m]
\ea

This fact assures that the singular values of $M_L$ and $M_R$ are equal. The argument follows from the trace Cayley Hamilton theorem \cite{Hong_Hao_2008} which states that for a complex $n\times n$ matrix $A$, the characteristic function
can be written as
\ba
    \chi_A (\lambda)
    &= \det(\lambda \mathds{1}-A) \nn\\
    &= \lambda^n+P_1 \lambda^{n-1} +P_2  \lambda^{n-2}+\cdots +P_{n-1}\lambda + P_n
\ea
where 
\ba
    P_k = 
    \sum_{(p_1,p_2,\ldots,p_k)\in S_k} \
    \prod_{m=1}^k \ 
    \frac{1}{p_m!} \qty(\frac{-\Tr[A^m]}{m})^{\!p_m}
\ea
and $S_k=\{(p_1,p_2,\ldots, p_k)\}$ is the set of non-negative integer solutions to the equation $\sum_{r=1}^k rp_r=k$.
When $\Tr[(M_L^{\dagger}M_L)^k]=\Tr[(M_R^{\dagger}M_R)^k]$ $\forall \ k\in \mathbb{N}$, then the singular values of $M_L^{\dagger}M_L$ and $M_R^{\dagger}M_R$ must be equal.
These singular values are the squares of the singular values of $M_L$, and $M_R$, which themselves are non negative, hence the singular values of $M_L$ and $M_R$ are equal.



\section{Reformulation of the current in terms of the Choi-Jamiolkowski state}\label{sec:CJS}

It is shown that the information current can be rewritten in terms of the difference in \Renyitwo entropies of the inner product of the Choi-Jamiolkowski state (CJS) associated with $M_L^\dagger M_L$ and $M_R^\dagger M_R$, respectively.
We introduce $\{\tilde M_{\beta_\mu}\}_\mu$, $\beta\in\{L,R\}$, as the set of (left or right partitioned) non-vectorized local tensor of the MPO corresponding to $M_\beta = \sum_\mu \tilde M_{\beta_\mu} \otimes \tilde M_{\beta_\mu}^*$, see Fig.~\ref{fig:MLMR_tilde}.
(Note that repeated indices are summed over, here and in subsequent figures in this section.)
\bfig
    \centering
    \includegraphics[width=\columnwidth]{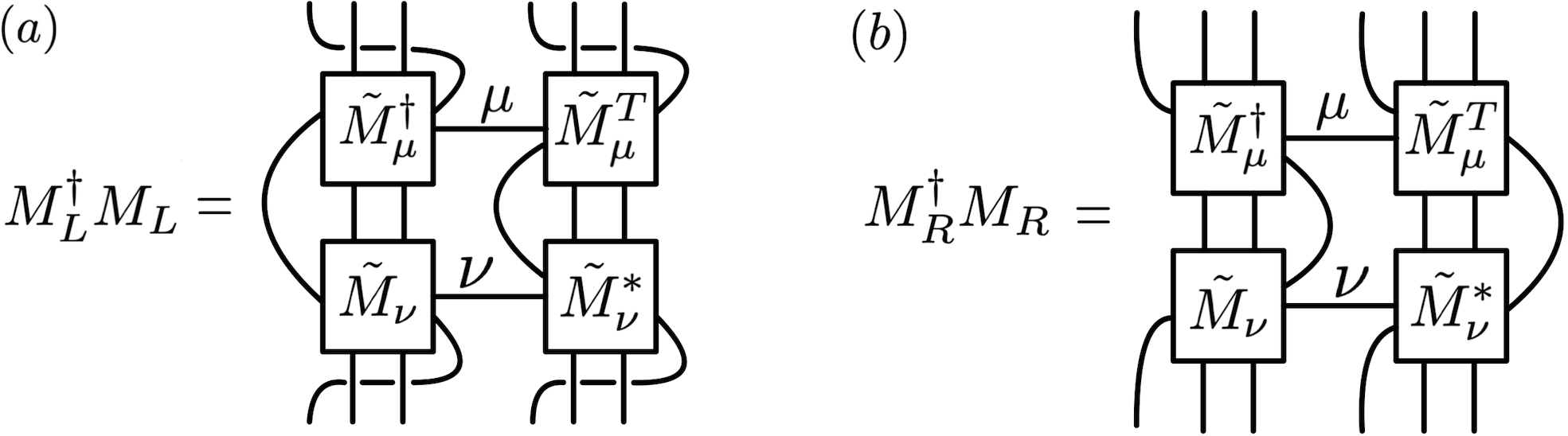}
    \caption{Tensor network description of (a) $M_L^\dagger M_L$, and (b) $M_R^\dagger M_R$, both rewritten in terms of the non-vectorized operators $\tilde M_\beta$; $M_\beta^\dagger M_\beta = \qty(\sum_{\mu} \tilde M_{\beta_\mu}^\dagger \otimes \tilde M_{\beta_\mu}^T )
    \qty(\sum_{\nu} \tilde M_{\beta_\nu} \otimes \tilde M_{\beta_\nu}^* )$.
    The subscript $\beta\in\{L,R\}$ in the tensor labels is omitted for clarity, as the associated left or right partitioning is given by the regrouping of virtual indices as defined in Fig.~\ref{fig:MLandMR}.
    }
    \label{fig:MLMR_tilde}
\efig

Given the map
\ba
    \epsilon_\beta(\rho)
    = \sum_{\mu,\nu} \qty( \tilde M_{\beta_\mu}^\dagger \tilde M_{\beta_{\nu}} ) \ \rho \ \qty( \tilde M_{\beta_{\nu}}^\dagger \tilde M_{\beta_{\mu}} ),
    \label{eq:epsilon}
\ea
the CJS is defined by
\ba
    \rho_\beta = \qty(\mathds{1}_{1,2,3}\otimes \epsilon_{\beta_{4,5,6}}) \qty(\rho_{\ket{\Phi^+}})
    \label{eq:CJS}
\ea
acting on
\ba
    \rho_{\ket{\Phi^+}} = \dyad{\Phi_d^+}_{1,6}\otimes\dyad{\Phi_d^+}_{2,5}\otimes\dyad{\Phi_D^+}_{3,4},
\ea
where $\ket{\Phi_d^+}=\frac{1}{\sqrt{d}}\sum_{j=0}^{d-1}\ket{j}\!\ket{j}$ are maximally entangled quantum states of two $d$-dimensional qudits; see Fig.~\ref{fig:CJS}. $\ket{\Phi_2^+}$ would represent a Bell pair of two qubits in case of $d=2$. Considering a QCA with a two-cell neighborhood, the two qudits on sites 5 and 6 represent the physical input states of the QCA on which the MPO, or here the map $\epsilon$, acts on. They are $d$-dimensional and are each maximally entangled with the qudit at site 1 or 2, respectively.
Since the local tensors of the MPO have a virtual degree of freedom of bond dimension $D$, see Fig.~\ref{fig:MLandMR}, we have introduced the $D$-dimensional qudit pair $\dyad{\Phi_D^+}_{3,4}$ on which the virtual degree of freedom of $\epsilon_\beta$ acts on.
\bfig
    \centering
    \includegraphics[width=\columnwidth]{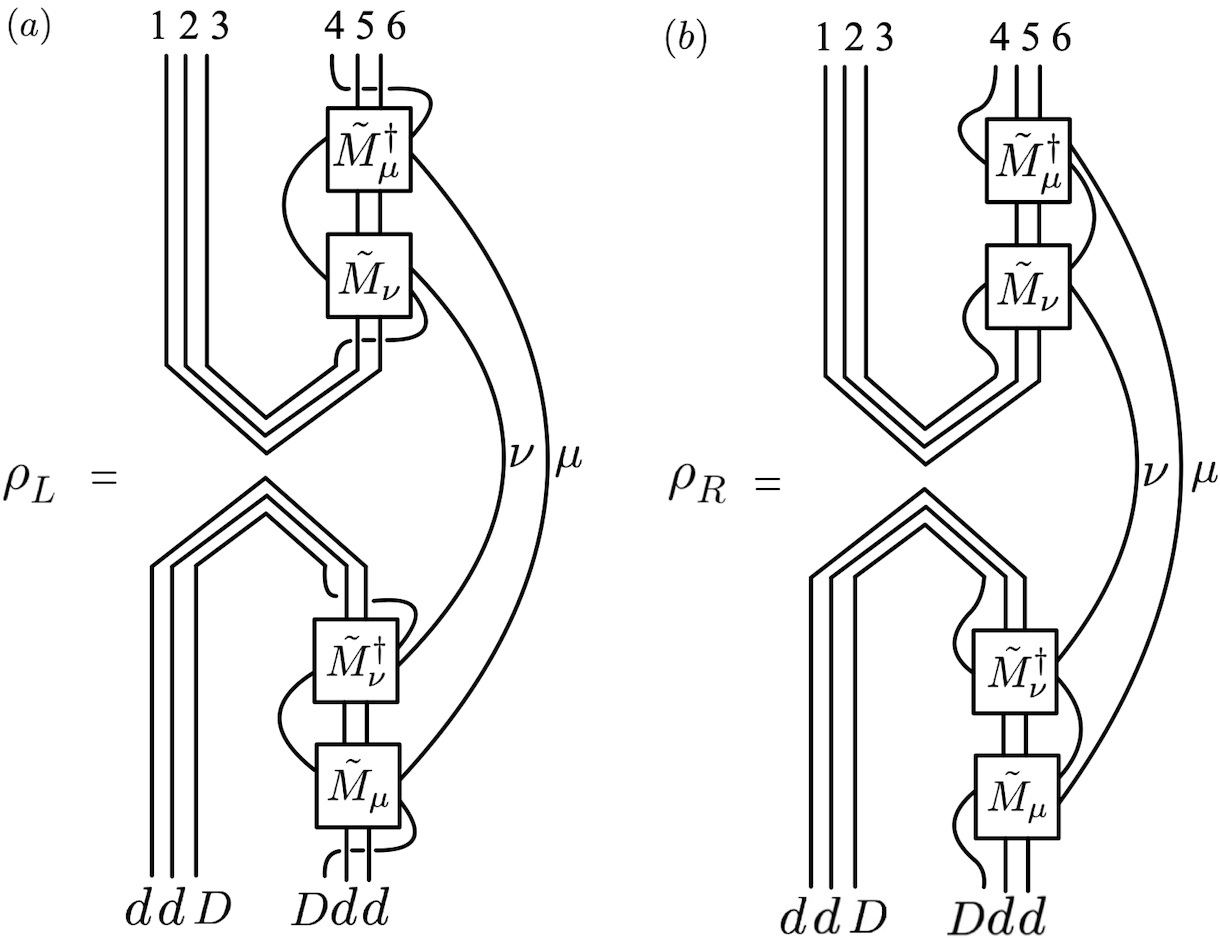}
    \caption{Tensor network description of the CJS associated with the quantum channels (a) $\epsilon_L(\rho)$ and (b) $\epsilon_R(\rho)$ shown in Eq.~\eqref{eq:epsilon}. $d$ is the physical dimension of each input state, while $D$ represents the bond dimension of the local MPO tensors $\tilde M_\mu$. 
    }
    \label{fig:CJS}
\efig
\bfig
    \centering
    \includegraphics[width=\columnwidth]{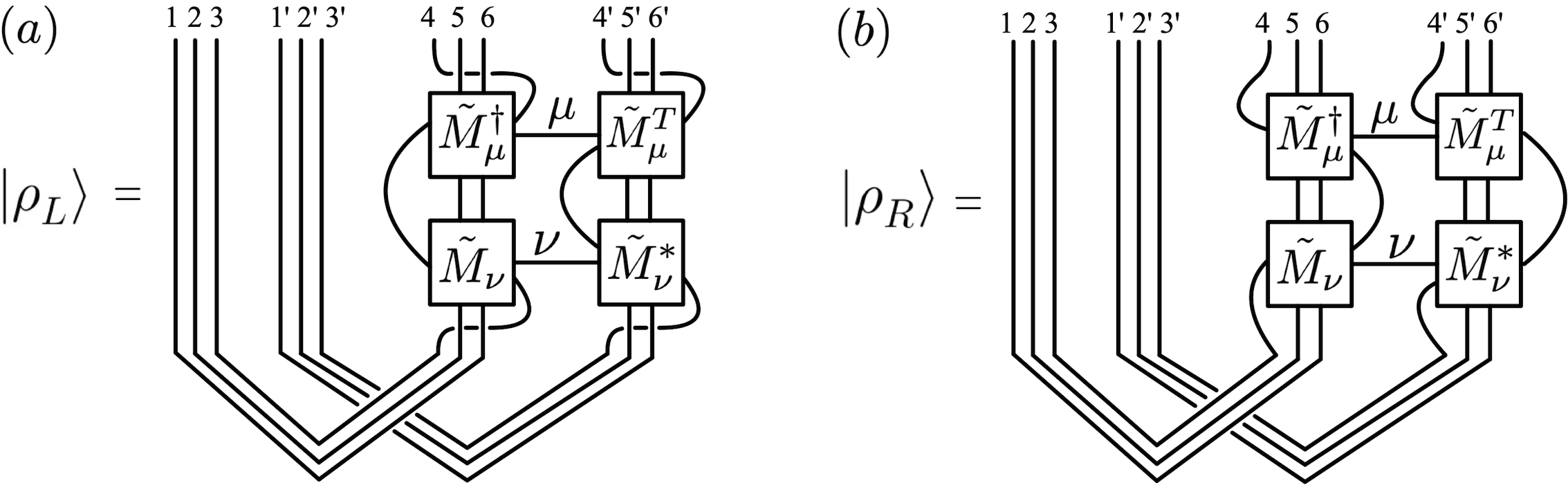}
    \caption{Vectorized version of the CJS of (a) $M_L^\dagger M_L$ and (b) $M_R^\dagger M_R$.
    }
    \label{fig:CJS_vectorized}
\efig

Using the \Renyitwo entropy $S_2(\rho) =-\log \Tr [\rho^2]$ and the identity $\Tr[\rho^2]=\braket{\rho}{\rho}$, where $\ket{\rho}$ is the vectorized state of $\rho$, see Fig.~\ref{fig:CJS_vectorized}, the current in Eq.~\eqref{eq:current} can then be rewritten in terms of the inner product of the CJS in Eq.~\eqref{eq:CJS}:
\ba
    I 
    = \frac{1}{2}\log\qty(\frac{\Tr[\rho_L^2]}{\Tr[\rho_R^2]})
    = \frac{1}{2}\log\qty(\frac{\braket{\rho_L}}{\braket{\rho_R}}),
\ea
as shown in Fig.~\ref{fig:CJS_current}.
\bfig
    \centering
    \includegraphics[width=\columnwidth]{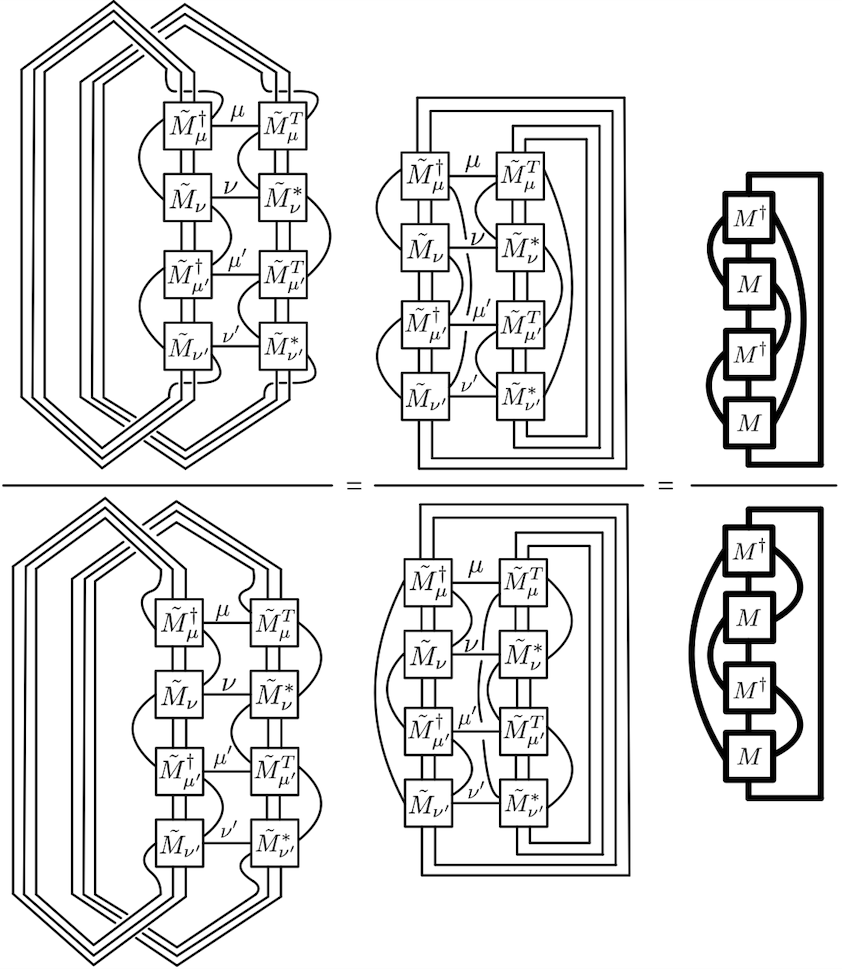}
    \caption{Argument of the current in Fig.~\ref{fig:current} rewritten in terms of the inner products of the CJS of $M_L^\dagger M_L$ and $M_R^\dagger M_R$, respectively, shown in Fig.~\ref{fig:CJS} and \ref{fig:CJS_vectorized}.
    }
    \label{fig:CJS_current}
\efig
State tomography of $\ket{\rho_\beta}$ could then determine the current.

\section{Proof of the invariance of \texorpdfstring{$I$}{TEXT} under local unitary conjugation of the gates \texorpdfstring{$V$}{TEXT} and \texorpdfstring{$W$}{TEXT}} \label{AppUnitary}
    
    It is derived that $I$ is invariant under conjugation of the local gates $V$ and $W$ with (the same) one-site unitaries $U$; writing $I\,(V,W)=I\,((U^\dagger\otimes U^\dagger) V (U\otimes U),(U^\dagger\otimes U^\dagger) W (U\otimes U))$.
    
    The invariance follows straight forward from the definition of a unitary matrix, $U U^\dagger=\mathds{1}$, and the cyclic property of the trace:
    \ba
        \Tr[\l(M_L^\dagger M_L\r)^{\!2}] \nn\\
        =
        \Tr
        \sum_{a,b,c,d=1}^\chi 
        &(U\otimes U)^\dagger\ \l(A_a^\dagger\otimes B_b^\dagger\r)\ (U\otimes U)
        \nn\\&\times
        (U\otimes U)^\dagger\  W^\dagger W\ (U\otimes U)
        &\nn\\&\times
        (U\otimes U)^\dagger\ \l(A_a\otimes B_c\r)\  (U\otimes U)
        \nn\\&\times
        (U\otimes U)^\dagger\ \l(A_d^\dagger\otimes B_c^\dagger\r)\ (U\otimes U)
        \nn\\&\times
        (U\otimes U)^\dagger\ W^\dagger W \ (U\otimes U)
        \nn\\&\times
        (U\otimes U)^\dagger\ \l(A_d\otimes B_b\r)\ (U\otimes U)
        \nn\\
        =\Tr
        \sum_{a,b,c,d=1}^\chi 
        &\l(A_a^\dagger\otimes B_b^\dagger\r)
        W^\dagger W
        \l(A_a\otimes B_c\r)
        \nn\\&\times
        \l(A_d^\dagger\otimes B_c^\dagger\r)
        W^\dagger W
        \l(A_d\otimes B_b\r),
    \label{eq:invarianceunderconjugation}
    \ea
    and analogously for  $\Tr[\l(M_R^\dagger M_{_R}\r)^{\!2}]$.

\section{Derivation of the current for the reset-swap QCA} \label{AppResetswap}

Here, the current for the reset-swap QCA is derived as a function of the reset gate $R$ and the set of Pauli operators $\{\sigma^a\}_{a=0}^3$. It is further shown that the current for this map does not change if one would add a local unitary $U$; i.e.~$I$ is invariant under ``transformations" $R\rarrow R U$, $R\rarrow U R$, or $R\rarrow U^\dagger R U$.
\begin{figure}[h]
    \includegraphics[width=\columnwidth]{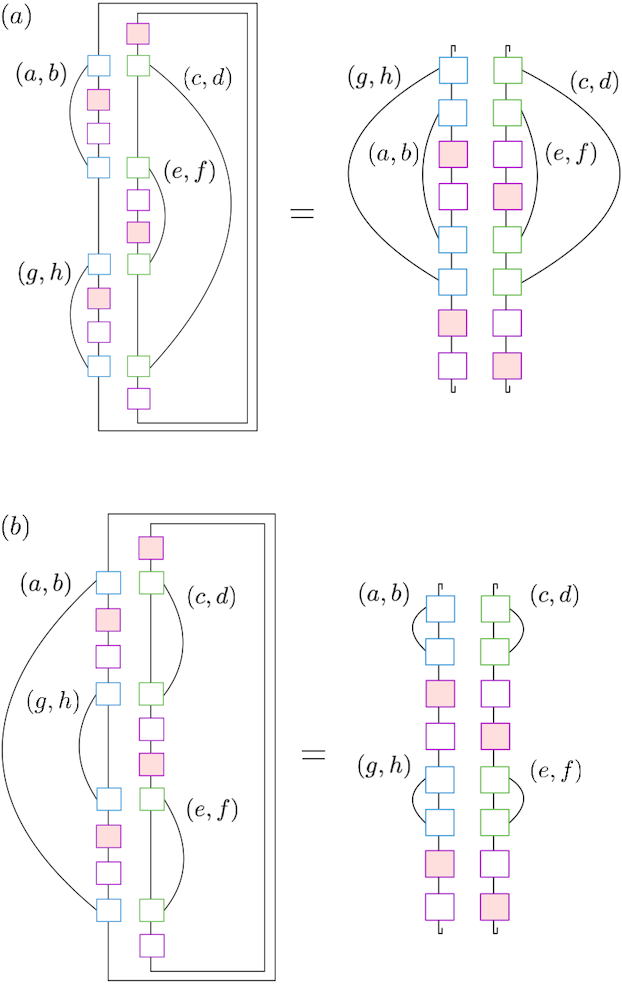}
    \caption{Tensor network description of 
    (a) $\Tr\!\l[\l(M_{L}^\dagger M_{L}\r)^{\!2}\r]$,
    (b) $\Tr\!\l[\l(M_{R}^\dagger M_{R}\r)^{\!2}\r]$,
    with added summation indices $(a,b),(c,d),(e,f),(g,h)$ used
    for the derivation of the current of the reset-swap QCA; see Eqs.~\eqref{eq:resetswapderivation1} and \eqref{eq:resetswapderivation2}.
    }
    \label{fig:resetswapcurrentderivation}
\efig

First, the traces of the second moments of $\l(M_{L,R}^\dagger M_{L,R}\r)$ are simplified as illustrated in Fig.~\ref{fig:resetswapcurrentderivation}: the left-hand sites in subfigures (a) and (b) are
\bs
\ba
    \Tr\!\l[\l(M_{L}^\dagger M_{L}\r)^{\!2}\r]
    &= 
        \frac{1}{2^8} \Tr\Big[
        \sum_{a,b,c,d,e,f,g,h=0}^3
        \nn\\&\qquad
        R_2^\dagger\,(\sigma^a\otimes\sigma^b)_1 \otimes (\sigma^c\otimes\sigma^d)_2 \,R_1^\dagger
        \nn\\&\qquad
        \times R_1\,(\sigma^a\otimes\sigma^b)_1 \otimes (\sigma^e\otimes\sigma^f)_2 \,R_2
        \nn\\&\qquad
        \times R_2^\dagger\,(\sigma^h\otimes\sigma^h)_1 \otimes (\sigma^e\otimes\sigma^f)_2 \,R_1^\dagger
        \nn\\&\qquad
        \times R_1\,(\sigma^g\otimes\sigma^h)_1 \otimes (\sigma^c\otimes\sigma^d)_2 \,R_2
        \Big],
    \\
    \Tr\!\l[\l(M_{R}^\dagger M_{R}\r)^{\!2}\r]
    &= 
        \frac{1}{2^8} \Tr\Big[
        \sum_{a,b,c,d,e,f,g,h=0}^3
        \nn\\&\qquad
        R_2^\dagger\,(\sigma^a\otimes\sigma^b)_1 \otimes (\sigma^c\otimes\sigma^d)_2 \,R_1^\dagger
        \nn\\&\qquad
        \times R_1\,(\sigma^g\otimes\sigma^h)_1 \otimes (\sigma^c\otimes\sigma^d)_2 \,R_2
        \nn\\&\qquad
        \times R_2^\dagger\,(\sigma^g\otimes\sigma^h)_1 \otimes (\sigma^e\otimes\sigma^f)_2 \,R_1^\dagger
        \nn\\&\qquad
        \times R_1\,(\sigma^a\otimes\sigma^b)_1 \otimes (\sigma^e\otimes\sigma^f)_2 \,R_2
        \Big],
\ea
\label{eq:resetswapderivation1}
\es
respectively.
Separating the operators acting on both, site 1 and site 2, leads to the expressions on the right-hand site of Fig.~\ref{fig:resetswapcurrentderivation}:
\bs
\ba
    \Tr\!\l[\l(M_{L}^\dagger M_{L}\r)^{\!2}\r]
    &= 
        \frac{1}{2^8} \Tr\Big[
        \sum_{a,b,g,h=0}^3 
        (\sigma^g\otimes\sigma^h)
        (\sigma^a\otimes\sigma^b) \,R^\dagger R
        \nn\\&\qquad\qquad\times
        (\sigma^a\otimes\sigma^b)
        (\sigma^g\otimes\sigma^h) \,R^\dagger R
        \Big]_1
        \nn\\&\qquad\times
        \Tr\Big[
        \sum_{c,d,e,f=0}^3 
        (\sigma^c\otimes\sigma^d)
        (\sigma^e\otimes\sigma^f) \,R R^\dagger
        \nn\\&\qquad\qquad\times
        (\sigma^e\otimes\sigma^f)
        (\sigma^c\otimes\sigma^d) \,R R^\dagger
        \Big]_2,
    \\
    \Tr\!\l[\l(M_{R}^\dagger M_{R}\r)^{\!2}\r]
    &= 
        \frac{1}{2^8} \Tr\Big[
        \sum_{a,b=0}^3 
        (\sigma^a\otimes\sigma^b)
        (\sigma^a\otimes\sigma^b) \,R^\dagger R
        \nn\\&\qquad\qquad\times
        \sum_{g,h=0}^3 
        (\sigma^g\otimes\sigma^h)
        (\sigma^g\otimes\sigma^h) \,R^\dagger R
        \Big]_1
        \nn\\&\qquad\times
        \Tr\Big[
        \sum_{c,d=0}^3 
        (\sigma^c\otimes\sigma^d)
        (\sigma^c\otimes\sigma^d) \,R R^\dagger
        \nn\\&\qquad\qquad\times
        \sum_{e,f=0}^3 
        (\sigma^e\otimes\sigma^f)
        (\sigma^e\otimes\sigma^f) \,R R^\dagger
        \Big]_2,
    \nn\\
    &= 
        2^8 \Tr[R^\dagger R R^\dagger R]_1 \, \Tr[R R^\dagger R R^\dagger]_2
    \nn\\
    &= 
        2^8 \l(\Tr[R^\dagger R R^\dagger R] \r)^2,
\ea
\label{eq:resetswapderivation2}
\es
where the subscripts 1 and 2 indicate the site all operators in the arguments of the respective traces are acting on. The last two simplifications follow from the identities presented in Fig.~\ref{fig:resetswapcurrent}(b) and (c).

The argument of the current is then in total:
\begin{equation}
    \begin{array}{lll}

    \frac{
    \Tr\!\l[\l(M_{L}^\dagger M_{L}\r)^{\!2}\r]
    }{
    \Tr\!\l[\l(M_{R}^\dagger M_{R}\r)^{\!2}\r]}
    &=&
    \frac{1}{
    \l(\Tr[R^\dagger R R^\dagger R] \r)^2
    }
        \Tr\Big[
            \sum_{a,b,g,h=0}^3 
            (\sigma^g\otimes\sigma^h)\\
            &&
            (\sigma^a\otimes\sigma^b) \,R^\dagger R
            (\sigma^a\otimes\sigma^b)
            (\sigma^g\otimes\sigma^h) \,R^\dagger R
            \Big]
    \\
        &\times&
        \Tr\Big[
            \sum_{c,d,e,f=0}^3 
            (\sigma^c\otimes\sigma^d)
            (\sigma^e\otimes\sigma^f) \,R R^\dagger\\
            &&
            (\sigma^e\otimes\sigma^f)
            (\sigma^c\otimes\sigma^d) \,R R^\dagger
            \Big].
\end{array}
\end{equation}

This expression does not change under composition with a unitary $U$,
where $U U^\dagger=\mathds{1}$, because the three terms
\ba
    (i)&\ 
        \Tr[R^\dagger R R^\dagger R] \nn\\
    (ii)&\
        \sum_{a,b=0}^3 (\sigma^a\otimes\sigma^b) R^\dagger R (\sigma^a\otimes\sigma^b) \nn\\
    (iii)&\
        \sum_{a,b=0}^3 (\sigma^a\otimes\sigma^b) RR^\dagger (\sigma^a\otimes\sigma^b) \nn
\ea
are invariant under the substitutions 
$R \rarrow R U$,
$R \rarrow U R$, and
$R \rarrow URU$.
The proof is shown below, where for each substitution the invariance of all three terms in $(i)$ to $(iii)$ is listed.
\bi
    \item $R \rarrow R U$, 
        \ba
            (i)&\ 
                \Tr[R^\dagger R R^\dagger R] \nn\\&\
                 = \Tr[R R^\dagger R R^\dagger] \nn\\&\
                 \rarrow \Tr[(RU) (U^\dagger R^\dagger) (RU) (U^\dagger R^\dagger)] \nn\\&\
                 = \Tr[R^\dagger R R^\dagger R] \nn\\
            (ii)&\
                \sum_{a,b=0}^3 (\sigma^a\otimes\sigma^b) R^\dagger R (\sigma^a\otimes\sigma^b) 
                \nn\\
                &\rarrow\sum_{a,b=0}^3 (\sigma^a\otimes\sigma^b) (U^\dagger R^\dagger) (R U) (\sigma^a\otimes\sigma^b)
                \nn\\
                &\qquad=\sum_{a,b=0}^3 (\sigma^a\otimes\sigma^b) R^\dagger R (\sigma^a\otimes\sigma^b),\nn\\
            (iii)&\
                 RR^\dagger 
                \rarrow (RU) (U^\dagger R^\dagger)
                = RR^\dagger
        \ea
    \item $R \rarrow U R$, 
        \ba
            (i)&\ 
                \Tr[R^\dagger R R^\dagger R] \nn\\&\
                 \rarrow \Tr[(R^\dagger U^\dagger) (UR) (R^\dagger U^\dagger) (UR)] \nn\\&\
                 = \Tr[R^\dagger RR^\dagger R] \nn\\
            (ii)&\
                R^\dagger R
                \rarrow (R^\dagger U^\dagger) (UR)
                = R^\dagger R \nn\\
            (iii)&\
                \sum_{a,b=0}^3 (\sigma^a\otimes\sigma^b) RR^\dagger (\sigma^a\otimes\sigma^b) 
                \nn\\
                &\rarrow\sum_{a,b=0}^3 (\sigma^a\otimes\sigma^b) (UR) (R^\dagger U^\dagger)  (\sigma^a\otimes\sigma^b)
                \nn\\
                &\qquad=\sum_{a,b=0}^3 (\sigma^a\otimes\sigma^b) RR^\dagger (\sigma^a\otimes\sigma^b),
        \ea
    \item $R \rarrow URU$, 
        \ba
            (i)&\
                \Tr[R^\dagger R R^\dagger R] \nn\\&\
                 \rarrow \Tr[(U^\dagger R^\dagger U^\dagger) (URU) (U^\dagger R^\dagger U^\dagger) (URU)] \nn\\&\
                 = \Tr[R^\dagger R R^\dagger R] \nn\\
            (ii)&\
                \sum_{a,b=0}^3 (\sigma^a\otimes\sigma^b) R^\dagger R (\sigma^a\otimes\sigma^b) 
                \nn\\
                &\rarrow\sum_{a,b=0}^3 (\sigma^a\otimes\sigma^b) (U^\dagger R^\dagger U^\dagger) (URU) (\sigma^a\otimes\sigma^b)
                \nn\\
                &\rarrow\sum_{a,b=0}^3 (\sigma^a\otimes\sigma^b) U^\dagger (R^\dagger R) U (\sigma^a\otimes\sigma^b)
                \nn\\
                &\qquad=\sum_{a,b=0}^3 (\sigma^a\otimes\sigma^b) R^\dagger R (\sigma^a\otimes\sigma^b) \nn\\
            (iii)&\
                \sum_{a,b=0}^3 (\sigma^a\otimes\sigma^b) R R^\dagger (\sigma^a\otimes\sigma^b) 
                \nn\\
                &\rarrow\sum_{a,b=0}^3 (\sigma^a\otimes\sigma^b) (U^\dagger R^\dagger U^\dagger) (URU) (\sigma^a\otimes\sigma^b)
                \nn\\
                &\qquad=\sum_{a,b=0}^3 (\sigma^a\otimes\sigma^b) U^\dagger (R^\dagger R) U (\sigma^a\otimes\sigma^b)
                \nn\\
                &\qquad=\sum_{a,b=0}^3 (\sigma^a\otimes\sigma^b) R^\dagger R (\sigma^a\otimes\sigma^b),
        \ea
\ei
Note that in Eqs.~$(i)$, the unitary condition is used, $U^\dagger U = \mathds{1} = UU^\dagger$,
while in Eqs.~$(ii)$, the unitary $U$ does not change the result, because it acts as a change of basis on the constituent evolution operators $\{\sigma^a\}_{a=0}^3$ of the swap operators.

The current is therefore invariant under composition with a unitary finite-depth circuit, and independent of the ordering of the composition with the reset gate.

Nonetheless, an additional local unitary gate would change the current if the QCA is coarse-grained and composed of two or more single time steps.
The proof is captured by below arguments of the current under composition in Eqs.~\eqref{eq:unitaryresetswapcompositionML} to \eqref{eq:unitaryresetswapcomposition}, and
pictured in Fig.~\ref{fig:resetswap3compositionwithunitary}.
\ba
    &\Tr\l[\l((M^n)^{\otimes^n}\r)_{\!_L}^\dagger \l((M^n)^{\otimes^n}\r)_{\!_L}\r]\nn\\
    &=\Tr\Big[
            \sum_{a,b,g,h=0}^3 
            (\sigma^g\otimes\sigma^h)(\sigma^a\otimes\sigma^b) 
            \,\l(U^\dagger R^\dagger\r)^{2n-1}\nn\\
            &\qquad(R U)^{2n-1}
            (\sigma^a\otimes\sigma^b)
            (\sigma^g\otimes\sigma^h)
            \,\l(U^\dagger R^\dagger\r)^{2n-1} (R U)^{2n-1}            \Big]_1
        \nn\\ &\times
        \Tr\Big[
            \sum_{c,d,e,f=0}^3
            (\sigma^c\otimes\sigma^d)
            (\sigma^e\otimes\sigma^f)
            \,(R U)^{2n-1}\nn\\
            &\qquad\l(U^\dagger R^\dagger\r)^{2n-1}
            (\sigma^e\otimes\sigma^f)
            (\sigma^c\otimes\sigma^d)
            \,(R U)^{2n-1} \l(U^\dagger R^\dagger\r)^{2n-1}
            \Big]_2
        \nn\\
       &\times
        \Tr\l[\l((M^{n-1})^{\otimes^{n-1}}\r)_{\!_L}^\dagger \l((M^{n-1})^{\otimes^{n-1}}\r)_{\!_L}\r],
\label{eq:unitaryresetswapcompositionML}
\ea

\ba
    &\Tr\l[\l((M^n)^{\otimes^n}\r)_{\!_R}^\dagger \l((M^n)^{\otimes^n}\r)_{\!_R}\r]
        \nn\\&\qquad
        =
        \l(\Tr\l[ \l( (U^\dagger R^\dagger)^{2n-1} (R U)^{2n-1} \r) \r] \r)^2
        \nn\\
        &\qquad\qquad\times\Tr\l[\l((M^{n-1})^{\otimes^{n-1}}\r)_{\!_R}^\dagger \l((M^{n-1})^{\otimes^{n-1}}\r)_{\!_R}\r],
\label{eq:unitaryresetswapcompositionMR}
\ea

\ba
    &\frac
    {\Tr\l[\l((M^n)^{\otimes^n}\r)_{\!_L}^\dagger \l((M^n)^{\otimes^n}\r)_{\!_L}\r]}
    {\Tr\l[\l((M^n)^{\otimes^n}\r)_{\!_R}^\dagger \l((M^n)^{\otimes^n}\r)_{\!_R}\r]}
    \nn\\&=
        \frac{1}{
        \l(\Tr\l[ \l( (U^\dagger R^\dagger)^{2n-1} (R U)^{2n-1} \r) \r] \r)^2 
        }
        \nn\\&\quad\times
        \Tr\Big[
            \sum_{a,b,g,h=0}^3 
            (\sigma^g\otimes\sigma^h)
            (\sigma^a\otimes\sigma^b) 
            \,\l(U^\dagger R^\dagger\r)^{2n-1}(R U)^{2n-1}\nn\\
            & \qquad\times
            (\sigma^a\otimes\sigma^b)
            (\sigma^g\otimes\sigma^h)
            \,\l(U^\dagger R^\dagger\r)^{2n-1} (R U)^{2n-1} \Big]_1
        \nn\\&\quad\times
        \Tr\Big[
            \sum_{c,d,e,f=0}^3
            (\sigma^c\otimes\sigma^d)
            (\sigma^e\otimes\sigma^f)
            \,(R U)^{2n-1}
            (U^\dagger R^\dagger)^{2n-1}\nn\\
            &\qquad\times
            (\sigma^e\otimes\sigma^f)
            (\sigma^c\otimes\sigma^d)
            \,(R U)^{2n-1} \l(U^\dagger R^\dagger\r)^{2n-1}
            \Big]_2
        \nn\\&\quad\times
            \frac
                {\Tr\l[\l((M^{n-1})^{\otimes^{n-1}}\r)_{\!_L}^\dagger \l((M^{n-1})^{\otimes^{n-1}}\r)_{\!_L}\r]}
                {\Tr\l[\l((M^{n-1})^{\otimes^{n-1}}\r)_{\!_R}^\dagger \l((M^{n-1})^{\otimes^{n-1}}\r)_{\!_R}\r]}
\label{eq:unitaryresetswapcomposition}
\ea

\begin{figure*}
    \includegraphics[scale=.5]{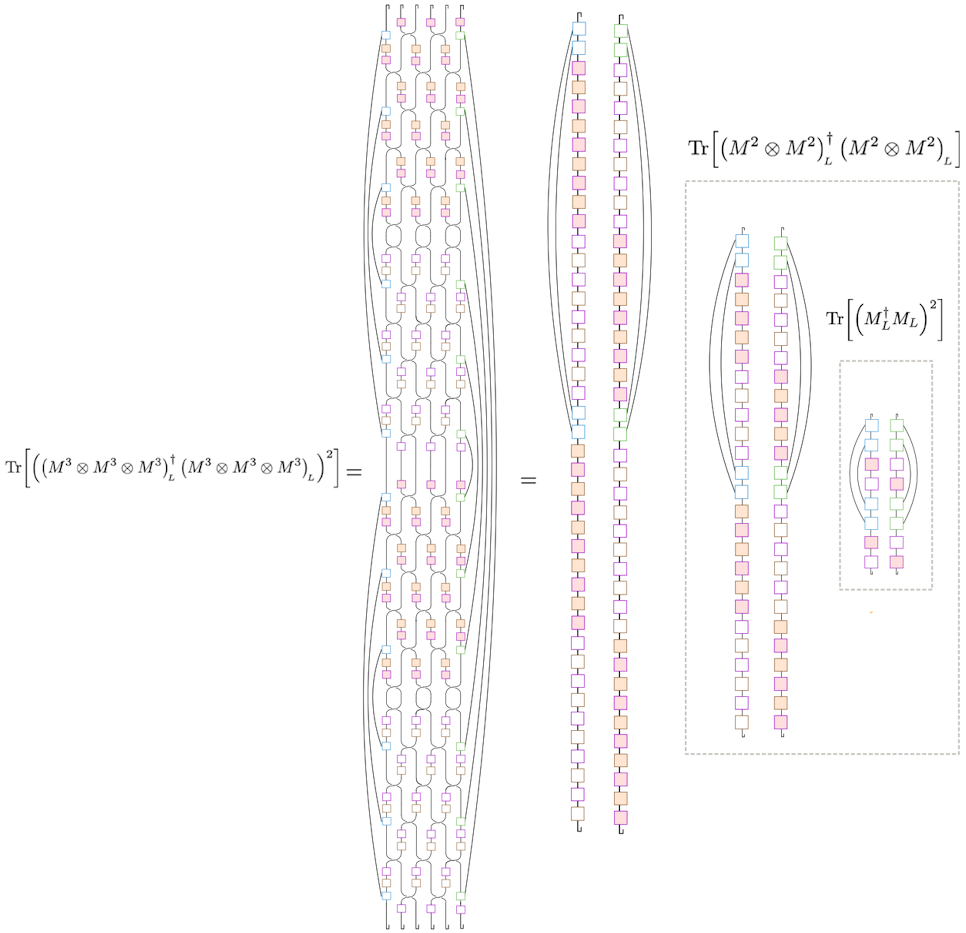}
    \caption{Tensor network description of 
    $
    \Tr\l[\l((M^n)^{\otimes^n}\r)_{\!_L}^\dagger \l((M^n)^{\otimes^n}\r)_{\!_L}\r]
    $
    in Eq.~\eqref{eq:unitaryresetswapcomposition}
    for the unitary-reset-swap QCA $(\SWAP_{1,2}\,R_1\,U_1)$ under composition of $n=3$ time steps.
    The same notation as is Figs.~\ref{fig:resetswap}, and \ref{fig:resetswapcurrent} is used, where the brown tensors represent the unitary superoperator, $(U\otimes U^*)$.
    }
    \label{fig:resetswap3compositionwithunitary}
\end{figure*}

\clearpage

\section{Derivation of \texorpdfstring{$W^\dagger W$}{TEXT} for the directed amplitude damping map}
\label{sec:Appamplitudedamping}
In this section it is shown that for the directed amplitude damping channel discussed in Sec.~\ref{sec:amplitudedamping}, $W^\dagger W$ is not separable when $p>0$.

The map is defined by the Kraus operators
\ba
    K_0=\m(1 & 0 & 0 & 0 \\0 & 1 & 0 & 0 \\0 & 0 & \sqrt{1-p} & 0 \\0 & 0 & 0 & 1), \
    K_1=\m(0 & 0 & \sqrt{p} & 0 \\0 & 0 & 0 & 0 \\0 & 0 & 0 & 0 \\0 & 0 & 0 & 0),
\ea
which determine the vectorized transfer matrix:
\ba
    W
    &= K_0 \otimes K_0 + K_1 \otimes K_1 \nn\\
    &= (
        \dyad{0}\otimes\dyad{0}
        + \dyad{0}\otimes\dyad{1} \nn\\
    &\qquad
        + \sqrt{1-p} \dyad{1}\otimes\dyad{0}
        + \dyad{1}\otimes\dyad{1}) \nn\\
    &\quad
    \otimes(
        \dyad{0}\otimes\dyad{0}
        + \dyad{0}\otimes\dyad{1} \nn\\
    &\qquad
        + \sqrt{1-p} \dyad{1}\otimes\dyad{0}
        + \dyad{1}\otimes\dyad{1}) \nn\\
    &\quad
        + p\,
            (\dyad{0}{1}\otimes\dyad{0})
            \otimes
            (\dyad{0}{1}\otimes\dyad{0}).
\ea
Applying the basis-change transformation from Eq.~\eqref{eq:Wbasischange},
\ba
    W
    \rarrow
    (\mathds{1}\otimes\hat\Sigma\otimes \mathds{1}) \
    W \
    (\mathds{1}\otimes\hat\Sigma\otimes \mathds{1}),
\ea
rearranges the order of subsystems in the tensor product, such that the operators can be combined which act on the same physical site, indicated by subscripts 1 and 2:
\ba
    W &=
    \dyad{00}_1\otimes
    \mathds{1}_2+
    \dyad{01}_1\otimes
    (\sqrt{1-p}\dyad{00}\nn\\
    &\quad+\dyad{01}+\sqrt{1-p}\dyad{10}+\dyad{11})_2 \nn\\
    &\quad+
    \dyad{10}_1\otimes
    (\sqrt{1-p}\dyad{00}+\sqrt{1-p}\dyad{01}\nn\\
    &\quad+\dyad{10}+\dyad{11})_2+
    \dyad{11}_1\nn\\
    &\quad\otimes
    ((1-p)\dyad{00}+\sqrt{1-p}\dyad{01} \nn\\
    &\quad+\sqrt{1-p}\dyad{10}
    +\dyad{11})_2\nn\\
    &\quad+
    p\dyad{00}{11}_1\otimes
    \dyad{00}_2.
    \label{eq:W}
\ea

Now one can write
\ba
W^{\dagger}W=\sum_{r,s=1}^{16} c_{r,s} P(r)_1\otimes P(s)_2,
\ea
where $\{P(a)\}_{a=1}^{16}$ is an orthonormal basis for two qubits. By explicit calculation one finds that for $p>0$, the matrix $c$ has four singular values whereas for $p=0$ there is only one as expected for the unitary case. Hence $W^{\dagger}W$ is not separable for $p>0$.

\clearpage

\end{document}

%% file: config/includes.tex
\usepackage{physics}
\usepackage{amssymb}
\usepackage{dsfont}
\usepackage{tabu,bm}
\usepackage{hyperref}
\usepackage{graphicx}
\usepackage{capt-of}
\usepackage{enumitem,cleveref}
\usepackage{centernot}
\usepackage{pifont} 
\usepackage[dvipsnames,table,x11names]{xcolor}
\DeclareGraphicsExtensions{.pdf,.jpeg,.jpg, .png, .eps, .tiff}


%% file: config/defs.tex
\newcommand{\ind}{{\rm ind}}
\newcommand{\Rank}{{\rm Rank}}
\newcommand{\Renyi}{R\'{e}nyi\ }
\newcommand{\Renyitwo}{R\'{e}nyi-2\ }
\newcommand{\SWAP}{\text{SWAP}}

\renewcommand{\epsilon}{\varepsilon}

\newcommand{\rarrow}{\rightarrow}
\renewcommand{\l}{\left}
\renewcommand{\r}{\right}

\newcommand{\tx}{\text}

\newcommand{\m}{\mqty}
\newcommand{\nn}{\nonumber}

\newcommand{\rom}[1]{\uppercase\expandafter{\romannumeral #1\relax}}

\newcommand{\cmark}{\ding{51}}
\newcommand{\xmark}{\ding{55}}

\def\ba#1\ea{\begin{align}#1\end{align}}
\def\bs{\begin{subequations}}
\def\es{\end{subequations}}
\def\beq{\begin{equation}}
\def\eeq{\end{equation}}
\def\bfig{\begin{figure}[h]}
\def\efig{\end{figure}}
\def\bsfig{\begin{subfigure}[b]}
\def\esfig{\end{subfigure}}
\def\bi{\begin{itemize}}
\def\ei{\end{itemize}}